\documentclass[12pt,a4paper]{article}
\usepackage{graphicx}
\usepackage{dcolumn}
\usepackage{bm}
\usepackage{amsmath,amsfonts,amssymb}
\usepackage{slashed}
\usepackage{braket,xcolor}
\usepackage{verbatim}
\usepackage{subcaption}
\usepackage{multirow}
\usepackage{amsfonts,mathtools,resizegather}
\usepackage[utf8]{inputenc}
\usepackage{caption,jheppub}
\DeclareMathOperator{\csch}{csch}

\title{\boldmath Partial islands and subregion complexity in geometric secret-sharing model}
\author[a]{Aranya Bhattacharya,}
\author[b]{Arpan Bhattacharyya,}
\author[a]{Pratik Nandy,}
\author[c]{Ayan K. Patra}

\affiliation[a]{Centre for High Energy Physics, Indian Institute of Science,\\ C.V. Raman Avenue, Bangalore 560012, India.}
\affiliation[b]{\textit{Indian Institute of Technology, Gandhinagar, Gujarat-382355, India}}
\affiliation[c]{Theory Division, Saha Institute of Nuclear Physics, HBNI, 1/AF Bidhannagar, Kolkata 700064, India}
\emailAdd{aranyab@iisc.ac.in}
\emailAdd{abhattacharyya@iitgn.ac.in}
\emailAdd{pratiknandy@iisc.ac.in}
\emailAdd{ayan.patra@saha.ac.in}

\abstract{We compute the holographic subregion complexity of a radiation subsystem in a geometric secret-sharing model of Hawking radiation in the ``complexity = volume" proposal. The model is constructed using multiboundary wormhole geometries in AdS$_{3}$. The entanglement curve for secret-sharing captures a crossover between two minimal curves in the geometry apart from the usual eternal Page curve present for the complete radiation entanglement.  We compute the complexity dual to the secret-sharing minimal surfaces and study their ``time" evolution. When we have access to a small part of the radiation, the complexity shows a jump at the secret-sharing time larger than the Page time. Moreover, the minimal surfaces do not have access to the entire island region for this particular case. They can only access it partially. We describe this inaccessibility in the context of ``classical" Markov recovery.}

\begin{document}
\maketitle
\flushbottom
\section{Introduction}
In the last few years, the particular problem of reproducing the Page curve \cite{Page:1993wv,Page:2013dx} for the black hole-radiation system has been solved using the ideas of Quantum Extremal Surfaces (QES) \cite{Engelhardt:2014gca, Penington:2019npb,Almheiri:2019hni, Almheiri:2019psf}. These surfaces are generalized versions of the Ryu-Takayanagi (RT) \cite{Ryu:2006bv, Ryu:2006ef, Hubeny:2007xt} surfaces, which measure the entanglement entropy between complementary parts of a holographic CFT. Hence, the ideas of holography or the AdS/CFT correspondence have played a central part in solving the puzzle as mentioned above within the broader version of the information paradox. Due to the application of QES, completely new bulk regions, known as the \textit{entanglement islands}, appear in the entanglement wedge of the radiation subsystem starting from a certain time (known as the Page time) of the evolution. Emergence of these islands results in reproducing the correct Page curve \cite{Page:1993wv,Page:2013dx} for both evaporating as well as the eternal black holes in AdS. In recent times, the Page curve has been explored in various contexts \cite{Penington:2019kki, Almheiri:2019psy,Hashimoto:2020cas, Anegawa:2020ezn, Alishahiha:2020qza, Gautason:2020tmk, Hartman:2020swn, Hollowood:2020cou, Geng:2020fxl, Geng:2020qvw, Li:2020ceg, Chandrasekaran:2020qtn, Akers:2019nfi, Balasubramanian:2020hfs, Hartman:2020khs, Balasubramanian:2020coy, Akal:2020twv, Akal:2021foz, Kawabata:2021hac, Kawabata:2021vyo, Ling:2020laa, Chen:2020jvn,Chen:2019uhq,Manu:2020tty, Chen:2020uac, Chen:2020hmv, Hollowood:2021nlo, Ghosh:2021axl, Chu:2021gdb, Caceres:2021fuw, Geng:2021hlu, Ahn:2021chg, Krishnan:2020oun, Hollowood:2021wkw, Li:2021dmf, Chen:2019iro, Liu:2020gnp, Akal:2020ujg, Krishnan:2020fer, Krishnan:2021faa, Bhattacharya:2020ymw, Basak:2020aaa, KumarBasak:2021rrx, Geng:2021wcq, Geng:2021iyq, Anderson:2021vof, Caceres:2020jcn, Reyes:2021npy, Neuenfeld:2021bsb, Matsuo:2020ypv, Azarnia:2021uch, Miyaji:2021lcq}. \par
The starting assumption is that the Hawking radiation is being absorbed by a non-gravitational bath coupled with the asymptotic boundary of the gravitational system containing the black hole. After that, one computes the entanglement entropy for a subregion for the radiation system by utilizing the $``$island$"$ formula \cite{Engelhardt:2014gca, Penington:2019npb, Almheiri:2019hni, Almheiri:2019psf, Almheiri:2020cfm}, 
	\begin{align} \label{eq1}
	S_{\mathrm{EE}}(\mathcal{R})=\mathrm{min} \bigg\{{\underset{\mathrm{islands}}{\mathrm{ext}}} \Big( S_{\rm{QFT}}(\mathcal{R} \cup \rm{islands})+\frac{\mathcal{A}(\partial \, (islands))}{4\, G_{N}}\Big) \bigg\}.
	\end{align}
{The equation (\ref{eq1}) takes into account both the entanglement entropy of quantum fields of radiation subregion $\mathcal{R}$ and the entanglement entropy of the gravitating subregions, termed as \textit{islands}}. For an evaporating black hole, the first term of (\ref{eq1}) dominates and the result matches with the Hawking’s evaluation of the entropy. But, at late times, island contribution i.e., the second term of (\ref{eq1}) dominates, and this reproduces the expected Page curve. However, the complete understanding of why these islands appear or what do they actually stand for is yet to be understood fully although some connections have been made with error correction and purification \cite{Bhattacharya:2020uun, Bhattacharya:2020ymw, Matsuo:2020ypv}. \par
Another crucially interesting direction is to track the time evolution of the black hole and radiation states. In doing so, one important quantity to study is the complexity of the states under scrutiny. Complexity measures the number of basic structural components needed to construct a given state \cite{Nielsen1133, Nielsen2006optimal}. Hence, applying the bulk proposals of complexity is one way to ask questions about the states we want to track. In a recent set of papers, the evolution of volume was studied. In \cite{Bhattacharya:2021jrn}, the authors found that the change of the preferred QES at Page time results in a jump (dip) in the volume corresponding to the radiation (black hole) degrees of freedom. Since these studies are completely based on the bulk proposal ``\textit{complexity $=$ subregion volume}", it is hard to understand precisely why this discontinuity appears in these pictures. However, in some of the models \cite{Bhattacharya:2020ymw}, the authors found this to be related to the holographic multipartite complexity of purification. The argument is that when the radiation subsystem suddenly gets access to a set of modes at Page time, which are spatially situated in the black hole interior, the partner modes which are already present in the radiation side get purified. From a gate counting perspective, therefore, while computing the complexity of all the modes in the radiation side, these purified modes should be treated differently as they do not need any auxiliary degrees of freedom to get purified. \par
In this paper, we focus on a slightly different problem. We consider only a part of the radiation system (say, half of the radiation system at each timestep) and see how does the information accessed by this region changes over time. We assume that as time goes on, the radiation is always stored in two different storage equally. Geometrically, this will become clear in the next section. This is in a similar spirit to \cite{Balasubramanian:2020hfs}. Here the authors found that even for half of the radiation subsystem, the corresponding QES goes through a shift at some point. This time is different from the Page time, and for the reasons that we explain later in this paper, we call this timescale the secret-sharing time $t_s$. As found in \cite{Balasubramanian:2020hfs}, the size of the entanglement wedge increases even for half of the radiation subsystem and gets access to the new set of modes from the black hole side at the secret-sharing time. We want to study how the corresponding volume (and therefore the sub-region complexity) changes at this point of shift. The overall goal is to have some idea about how parts of the radiation subsystem access the information and how it differs from the total radiation subsystem. In a way, this is also supposed to teach us (at least holographically) how the complexity of different parts of a mixed state behaves in comparison to the total mixed state. \par
As we find out, the subregion complexity of the half of the radiation subsystem also goes through a shift when the corresponding entanglement entropy shift happens. This shift, from the radiation point of view, is again similar to what was found in previous studies concerning the complete radiation subsystem making the phenomena of phase transition of complexity a bit more universal and applicable to a broader set of situations. However, the jump happens at the secret-sharing time, which is typically bigger than the Page time. \par
The rest of the paper is organised as follows. In section \ref{sec2}, we briefly review the model studied in \cite{Balasubramanian:2020hfs}, in which we do the computations. In section \ref{sec3}, we discuss the volume computations along with the plots, which will play the central role in the paper. We discuss the inaccessibility of certain regions in island in the context of ``classical" Markov recovery in section \ref{Markovapp}. Finally, in section \ref{secdiscuss}, we conclude with the implications of our study and try to understand why and how the results are plausible from the point of view of both information theory and the holographic model of the black hole-radiation system. 
\begin{figure}
	\centering
	\includegraphics[scale=0.4]{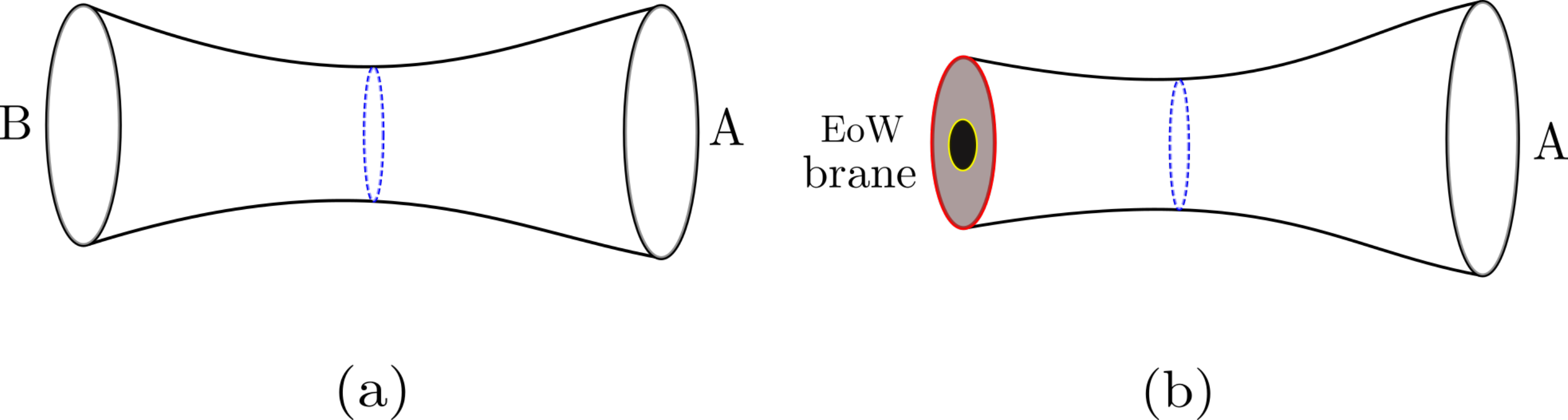}
	\caption{The inception geometry. (a) A two-boundary wormhole with asymptotic boundaries $A$ and $B$. The blue dotted circle denotes the bifurcate horizon. (b) The side $B$ is truncated and replaced by an EoW brane (red circle). The gray region is the inception disk containing the inception black hole (black) with its inception horizon (yellow circle).}
	\label{wormhole1}
\end{figure}
\section{Geometric secret-sharing model: a brief review}\label{sec2}
As we have mentioned previously, our study will be in continuation of the geometric secret-sharing model studied in \cite{Balasubramanian:2020hfs}. In this section, we briefly review the essential features of the model, which will help us to calculate the holographic subregion complexity. We also follow a treatment involving Killing vectors in some stages while reviewing the covering space depiction, which will be particularly relevant for the volume computations in section \ref{sec3}. The review part is important because it helps the reader to understand which volumes are to be computed and what precise parameter values are important in such computations. 

The secret-sharing model was primarily introduced to study the Page curve in $(2+1)$-dimensional AdS spacetime. The model essentially starts by considering a time slice of a pure eternal two-sided BTZ black hole and then cutting off one asymptotic side by putting an End-of-the-World (EoW) brane. This amounts to introducing a CFT on the brane itself, and one considers its holographic dual geometry which is termed as the inception geometry. Due to the entanglement with radiation, the brane CFT is thermal and dual to a black hole referred as the inception black hole. The inception geometry, which possesses a black hole with its intrinsic inception horizon, is purified by entanglement through another wormhole. This purifying auxiliary system is naturally identified with the external radiation, providing a realization of the $\mathrm{ER} = \mathrm{EPR}$ \cite{Maldacena:2013xja, Verlinde:2020upt}. The basic structure is shown in Fig.\ref{wormhole1}. Note that the AdS radius, the Newton's constant, and hence the central charge (of the dual CFTs) on two sides differ, which is the crux of the model. The following junction conditions glue the real side and inception side \cite{Balasubramanian:2020hfs}
\begin{align}
h_{ab} = h'_{ab},  ~~~~~~ \frac{1}{G_{N}}K_{ab} = \frac{1}{G'_{N}}K'_{ab},\label{2.1}
\end{align}
where $h_{ab}$ and $K_{ab}$ are the induced metric and extrinsic curvature at the junction, respectively. Here we use primed coordinates for the inception side, whereas the un-primed coordinates denote the real side of the geometry. The junction conditions are similar in spirit to Israel's junction conditions \cite{Israel:1966rt}, but there is a subtle difference - here, a convex surface is glued with another convex surface, whereas in Israel's condition, a concave surface is glued with a convex surface. For more details, we urge the interested reader to go through section 2 of \cite{Balasubramanian:2020hfs}. 
\begin{figure}
	\centering
	\includegraphics[scale=0.3]{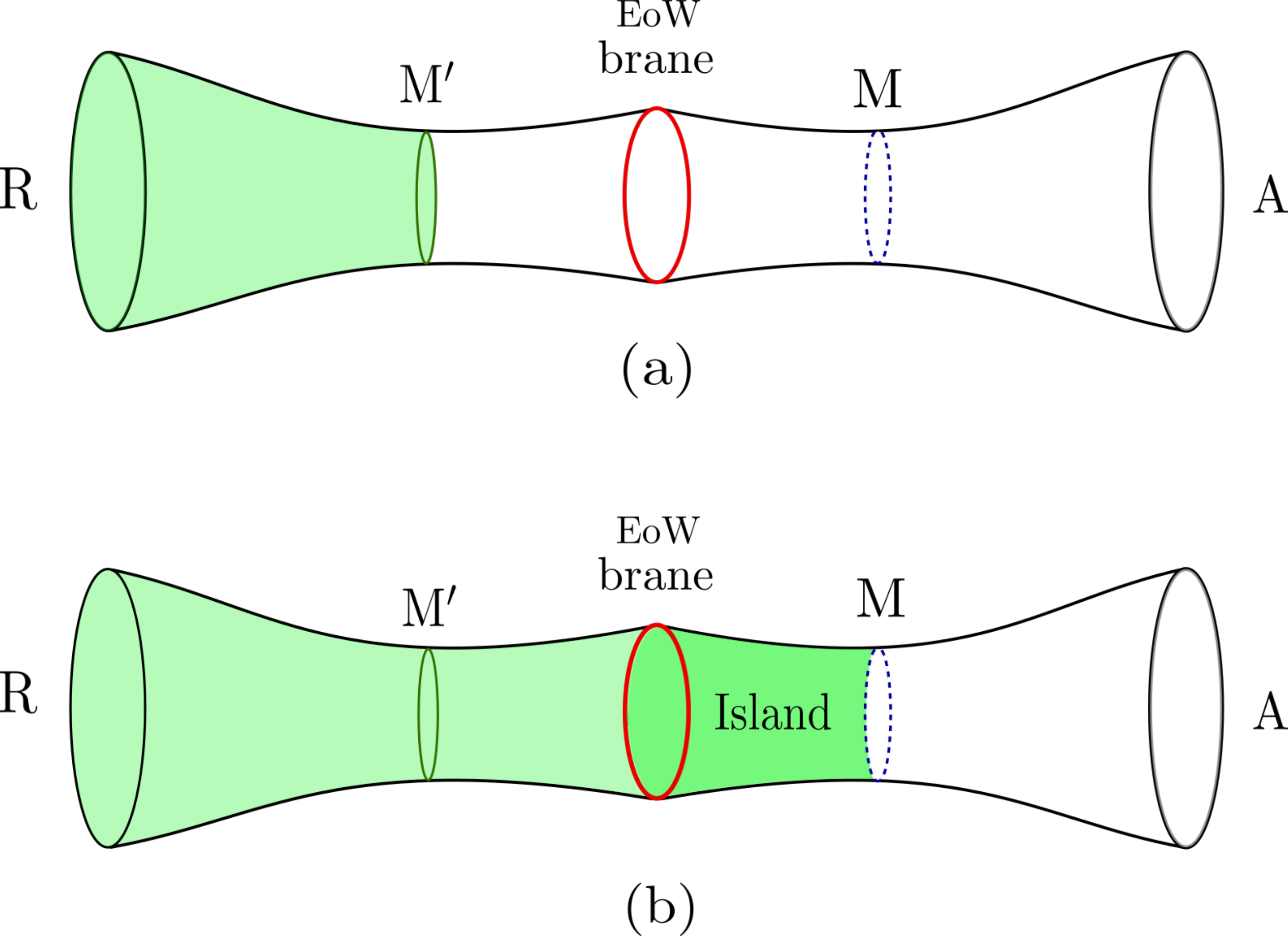}
	\caption{Entanglement wedge of radiation subsystem $R$ (a) before and (b) after Page time. (a) The green shaded region is the EW of $R$ as $M'$ is the preferred RT surface. (b) At Page time, the RT surface of $R$ changes to $M$, which goes through the EoW brane (marked as red) and contains part of the real black hole (the island region is shown by deep green). Here the blued dotted circle denotes the RT surface of the real black hole. The left side of the EoW brane is the inception side, whereas the right side of it is the real side.}
	\label{wormisl}
\end{figure}

We consider the Euclidean black hole geometry,
\begin{equation}
  ds^2=\frac{r^2-r_h^2}{\ell^2}d\tau^2+\frac{\ell^2 dr^2}{r^2-r_h^2}+r^2d\phi^2
  \label{eq2.2}
\end{equation}
where $r_h$ is the radius of the horizon. Inception side is also given by the same metric $\eqref{eq2.2}$ with horizon radius $r'_h$ and AdS radius $\ell'$. Using $\eqref{eq2.2}$, the junction conditions was already found in \cite{Balasubramanian:2020hfs}. We need two quantities from \cite{Balasubramanian:2020hfs} which are $r_t$ and $r_b$, to describe the evaporation protocol,
\begin{equation}
 r_t=\sqrt{\frac{\ell^2G_{N}^2 r'^2_h-\ell'^2G'^2_N r_h^2}{\ell^2G_N^2-\ell'^2G'^2_N}},~~~~~r_b=\sqrt{\frac{\ell^2r'^2_h-\ell'^2r_h^2}{\ell^2-\ell'^2}}.
\end{equation}
Here, $r_t$ is the location of the EoW on the upper half plane while at $r_b$ the brane trajectory satisfies $dr/d\tau=\infty$. The evaporation protocol is to gradually increase the radius $r_h'$ corresponding to the radiation subsystem. The Page transition is marked by the equality $r_h/G_N =r'_h/G'_N$, where $G_N$ and $G_N'$ are the Newton's constants from the real and inception side respectively. To see the transition before  $r'_h=r_h$, we require $G'_N < G_N$. We also fix $\ell$ and $G_N$ from the real side and allow $\ell'$ and $G_N'$ to vary from the inception side so that the ratio of central charges $\hat{c}= c/c'$ is fixed, where $c=3 \ell/2G_N$. In order to achieve this,
during the evaporation process, we vary the location of the EoW brane $r_t$ as,
\begin{equation}
    r_t=r_h+\alpha(r_h-r'_h),
\end{equation}
with $\alpha>0$. Let us consider the situation before Page time when ${r'_h}/{G'_N}<{r_h}/{G_N}$, and we focus on the radiation system $R$, which is situated in the inception side (see Fig.\ref{wormisl}). Before Page time, the RT surface of the region $R$ is the throat horizon $M'$, which has the minimum area. After the Page time, the RT surface of $R$ is given by the real black hole horizon $M$. Hence a part of the real black hole interior enters within the entanglement wedge of the radiation. This so-called ``island" part is shown in Fig.\ref{wormisl}$b$. The important point is that after the Page time, if we can access the full radiation, then we have the full information of the ``island" coming from the black hole interior.
\begin{figure}
	\centering
	\includegraphics[scale=0.47]{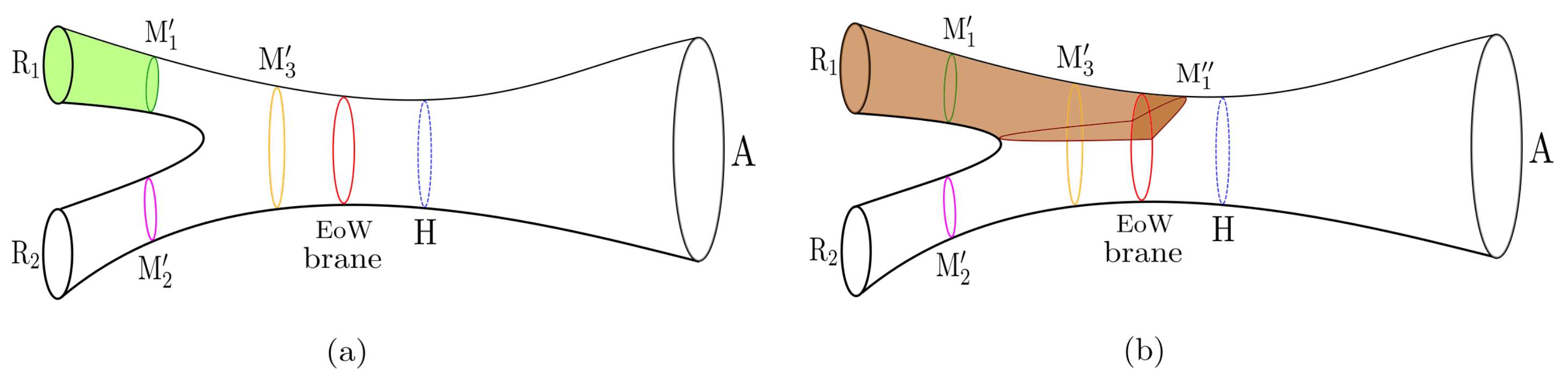}
	\caption{Entanglement wedge of radiation subsystem $R_1$ (a) before and (b) after Page time. (a) The green shaded region is the EW of $R_1$ as $M_1'$ is the preferred RT surface. (b) At Page time, the RT surface of $R_1$ changes to $M_1''$ (marked as brown), which goes through the EoW brane (red) and partially contains the interior of the real black hole. The EW of $R_1$ after Page time is shown by the brown shaded region, which includes a part of the island marked by the deeper brown colour. Here the blued dotted circle denotes the RT surface of the real black hole. $M_2'$ and $M_3'$  denote the RT surface of radiation subsystem $R_2$ and $R_1 \cup R_2$ before Page time respectively.}
	\label{wormhole5}
\end{figure}

On the other hand, if we only focus on a subsystem of radiation, the situation differs significantly. For example, let us consider the radiation subsystem $R_1$ before the Page time. Clearly the RT surface of $R_1$ is given by the throat horizon $M_1'$ (see Fig.\ref{wormhole5}). However, after a particular time (secret-sharing time), the RT surface changes significantly, it goes through the EoW brane (the fact that the RT surface can go through the EoW brane is a crucial assumption in this model, which was argued consistently in \cite{Balasubramanian:2020hfs}) and contains a part of the island, and thus given by the horizon $M_1''$. The key fact is that if we have access to the $R_1$ part only, we can never access the full ``island" region. We only have partial access.

Another important point is that the time at which the RT surface of $R_1$ changes is different and in fact, bigger than the Page time. In terms of the access to the partial islands, although the full radiation has access to the full ``island" region starting from Page time, only a part of that information can be accessed by the $R_1$ region after the secret-sharing time.  Even if one considers the union of partial island regions accessed by $R_1$ and $R_2$, a certain region within the full island remains inaccessible. Therefore, it is extremely crucial for $R_1$ and $R_2$ to communicate between themselves in an entangled manner (quantum channel) so that they can access to the ``\emph{island}$- 2 \,\times$\emph{partial island}" region. 

From the entanglement wedge (EW) point of view, if two observers are collecting the Hawking radiation separately at $R_1$ and $R_2$ exits, and they can communicate only through classical channels, a part of the black hole still remains \emph{secret} to both of them, because the union of EW of $R_1$ and $R_2$ is different than the full EW of $R_1 \cup R_2$, i.e., $\mathrm{EW}_{R_1} \cup \mathrm{EW}_{R_2} \neq \mathrm{EW}_{R_1 \cup R_2}$. A part of the interior of the black hole remains \emph{secret} to the outside observer. The interior region is fully accessible to the outsider only if a single observer has full access to both the $R_1$ and $R_2$ exists simultaneously, which means the full Hawking radiation is required to construct the full interior of the black hole. This is precisely the reason behind the name  ``secret-sharing"  of this particular model and the nomenclature of the new timescale.

In the next subsection, we discuss the procedure of the construction of the multiboundary wormhole with relevant computations.

\subsection{Covering space depiction of multiboundary wormholes}
\begin{figure}
	\centering
	\begin{subfigure}[b]{0.49\textwidth}
		\centering
		\includegraphics[width=\textwidth]{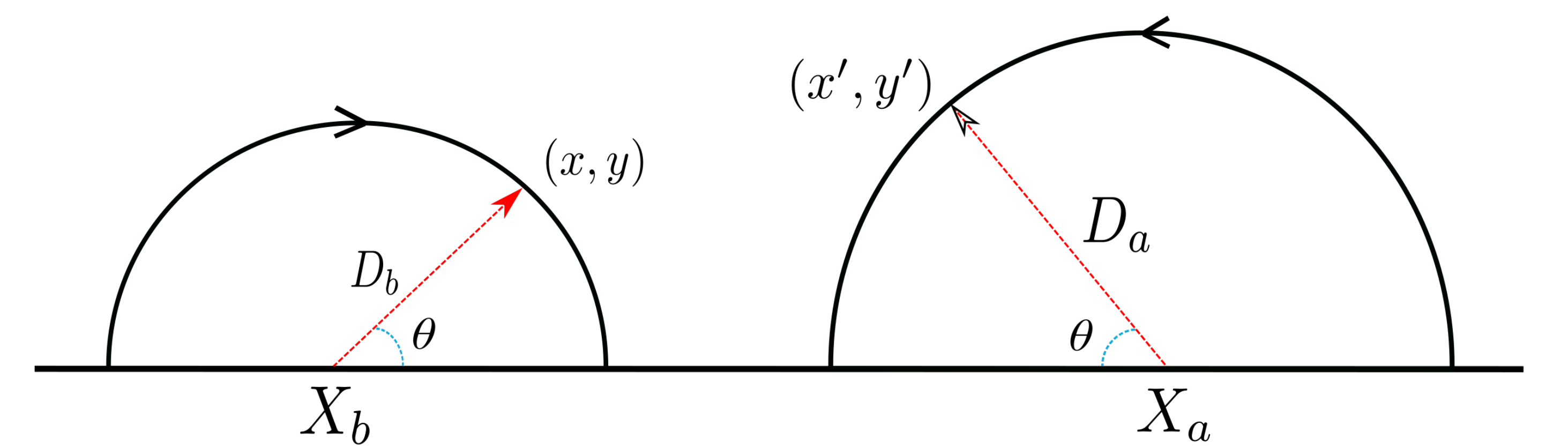}
		\caption{}
		\label{fig:mbw2}
		\end{subfigure}
		\hfill
	\begin{subfigure}[b]{0.5\textwidth}
		\centering
		\includegraphics[width=\textwidth]{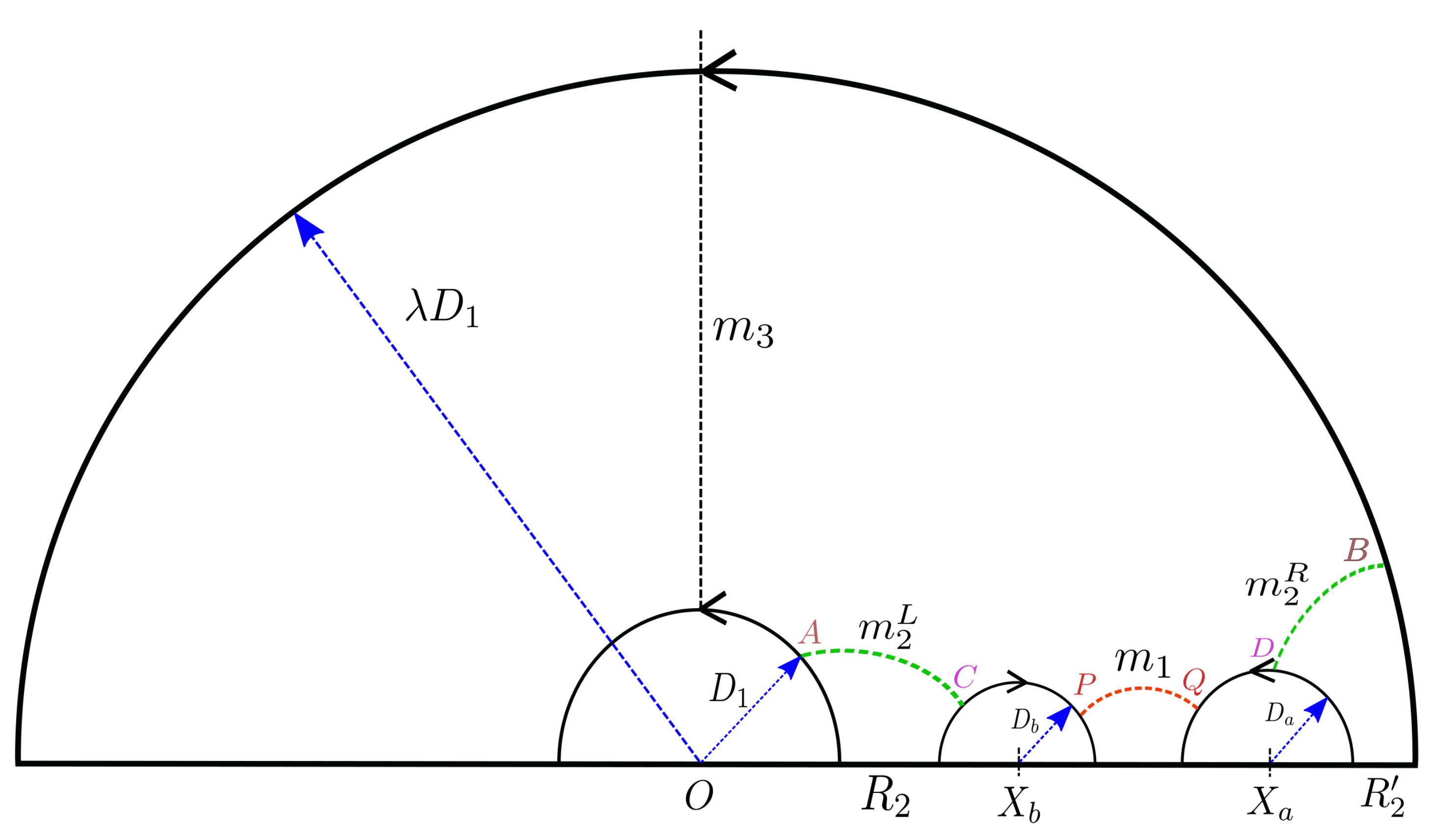}
		\caption{}
		\label{fig:mbw3}
		\end{subfigure}
		\caption{(a) The general orientation reversing isometry that maps the point $(x,y)$ on $g_b$ circle to a point $(x',y')$ on $g_a$ circle. (b) The covering space of the multiboundary wormhole. The notations are explained in the main text.}
		\label{fig:mbw23}
\end{figure}

In this section we describe the relation between the inception side and the covering space depiction of the multiboundary wormholes. The treatment of this section is different from that of \cite{Balasubramanian:2020hfs}. We follow the Killing vector approach developed in \cite{Caceres:2019giy} in writing down different moduli that play the role of the extremal surfaces in this model.
Static multiboundary wormholes can be constructed by taking a quotient of $t=0$ slice of AdS$_3$ which is a hyperbolic plane $\mathbb{H}$ with a discrete subgroup  $\Gamma$ of $\text{PSL} \, (2, \mathbb{R})$ known as Fuchsian group and then lifting the action of $\Gamma$ to the full AdS$_3$ \cite{Bhattacharyya:2016hbx}. The action of $\Gamma$ identifies pairs of geodesics on $t=0$ of AdS$_3$ to obtain $t=0$ slice of multiboundary wormholes. Then the covering space of $t=0$ slice of a multiboundary wormhole is just the quotient space of the hyperbolic plane or is $\mathbb{H}/\Gamma$. We take upper half-plane (UHP) metric as the $t=0$ slice of AdS$_3$. One of the geodesics in the UHP are boundary anchored semicircles, and we identify these semicircles to create our desired multiboundary wormhole geometries \cite{Balasubramanian:2014hda, Caceres:2019giy}.
To elaborate this point, we identify a pair of boundary anchored concentric semicircles on the UHP to create a $t=0$ slice of two-boundary wormhole. In a similar way we can construct a three-boundary wormhole by identifying a pair of concentric semicircles anchored at the boundary along with the identification of a pair semicircles in a reverse orientation way \cite{Caceres:2019giy} depicted in Fig.\ref{fig:mbw3}. We briefly demonstrate this reverse orientation identification in detail. Let us suppose that two circles are located at $X_b$ and $X_a$ with radius $D_b$ and $D_a$ respectively (see Fig.\ref{fig:mbw2}). The general orientation reversing isometry that maps a point $(x,y)$ on $g_b$ circle to a point $(x',y')$ on $g_a$ circle is,
\begin{align}
x' = X_a+\frac{D_a}{D_b}(X_b-x), ~~~~~~~
y' = \frac{D_a}{D_b}y\,.\label{tr0}
\end{align}
The transformations in \eqref{tr0} are finite transformations, and their corresponding infinitesimal transformations were found in \cite{Caceres:2019giy}. For the sake of completeness, we write down the Killing vector that corresponds to a infinitesimal transformation. The Killing vector is\cite{Caceres:2019giy},
\begin{equation}
\xi= a J_T+b J_D+c J_S\,.
\end{equation}
where $J_T$, $J_D$, and $J_S$ are the generators of translation, dilatation and special conformal transformation respectively \cite{Caceres:2019giy}. Let us define new parameters as $\bar{a}=-c, \bar{b}=-b/2c$ and $\bar{c}= \sqrt{b^2-4ac}/2c$.
In terms of these parameters we find the finite transformation to be
\begin{align}
z' = e^{\xi}z ~~~
\Rightarrow ~~~ z'= \bar{b}+\bar{c}\,\frac{(z-\bar{b})\cosh(\bar{a}\bar{c})+\bar{c}\sinh(\bar{a}\bar{c})}{(z-\bar{b})\sinh(\bar{a}\bar{c})+\bar{c}\cosh(\bar{a}\bar{c})}.
\end{align}
where $z=x+i y$, is a point on the hyperbolic plane $\mathbb{H}$.
We express the center and the radius of the identified semicircles in terms of these parameters
\begin{align}
    X_b = \bar{b}-\bar{c}\coth(\bar{a}\bar{c}), ~~~ X_a =\bar{b}+\bar{c}\coth(\bar{a}\bar{c}), ~~~ D_bD_a =  \bar{c}^2\csch^2(\bar{a}\bar{c}) \label{tran}
\end{align}
Finally, using \eqref{tran}, we find
\begin{equation}
D_b D_a=\left(\frac{X_a-X_b}{2}\right)^2-\bar{c}^2\,.
\end{equation}
{Wormhole horizons at $t=0$ slice become minimal periodic geodesics in the UHP.  Thus,  we look for minimal geodesics between any two identified semicircles.} To have a well defined fundamental domain shown in Fig.\ref{fig:mbw23}$b$ we require
\begin{equation}\label{domaincond}
D_1<X_b-D_b<X_b+D_b<X_a-D_a<X_a+D_a<\lambda D_1.
\end{equation}
Three boundary wormholes have three independent moduli $m_1$, $m_2$ and $m_3$.
The  horizon length $m_3$ is computed by evaluating the length of the vertical geodesic between two concentric semicircles
\begin{equation}
m_3=\ell'\int_{D_1}^{\lambda D_1}\frac{dy}{y}=\ell' \log\lambda\,.
\end{equation}
The procedure to find the moduli $m_1$ is the following: $i$) First, we take an arbitrary point on any of the two circles (say left) and its image point, which is determined by the relation \eqref{tr0} on the other circle (say right). $ii$) We find the length between these two points and minimize it with respect to their endpoints.
This procedure yields the two endpoints of the moduli $m_1$ as
\begin{eqnarray}
P&\equiv(P_x,P_y)= \left(X_b+\frac{D_b(D_b+D_a)}{X_a-X_b},\frac{D_b}{X_a-X_b}\sqrt{(X_a-X_b)^2-(D_b+D_a)^2}\right),\label{2.13}\\
Q&\equiv(Q_x,Q_y)=\left(X_a-\frac{D_a(D_b+D_a)}{X_a-X_b},\frac{D_a}{X_a-X_b}\sqrt{(X_a-X_b)^2-(D_b+D_a)^2}\right).\label{2.14}
\end{eqnarray} 
Hence, the length $m_1$ is
\begin{equation}
m_1=\ell'(\sinh^{-1}(\alpha)+\sinh^{-1}(\beta)),
\end{equation}
where
\begin{align}
    \alpha = \frac{(X_a-X_b)^2-2 D_a(D_b+D_a)}{2 D_a\sqrt{(X_a-X_b)^2-(D_b+D_a)^2}}, ~~~ \beta =\frac{(X_a-X_b)^2-2 D_b(D_b+D_a)}{2 D_b\sqrt{(X_a-X_b)^2-(D_b+D_a)^2}}\,.
\end{align}
Computing the horizon length $m_2$ is non-trivial as it is divided into two parts $m_2^L$ and $m_2^R$. Nevertheless, we follow the same procedure described above and find the length $m_2$ as
\begin{equation}
m_2=\ell'(\sinh^{-1}(\gamma)+\sinh^{-1}(\delta)),
\end{equation}
with
\begin{align}
    \gamma = \frac{(X_b-\lambda^{-1} X_a)^2-2D_b(D_b+\lambda^{-1}D_a)}{2D_b\sqrt{(X_b-\lambda^{-1} X_a)^2-(D_b+\lambda^{-1}D_a)^2}},~ \delta = \frac{(\lambda X_b- X_a)^2-2D_a(\lambda D_b+D_a)}{2D_a\sqrt{(\lambda X_b- X_a)^2-(\lambda D_b+D_a)^2}}.
\end{align}
\subsection{Islands and entanglement entropy}
\begin{figure}[t]
	\centering
	\includegraphics[scale=0.3]{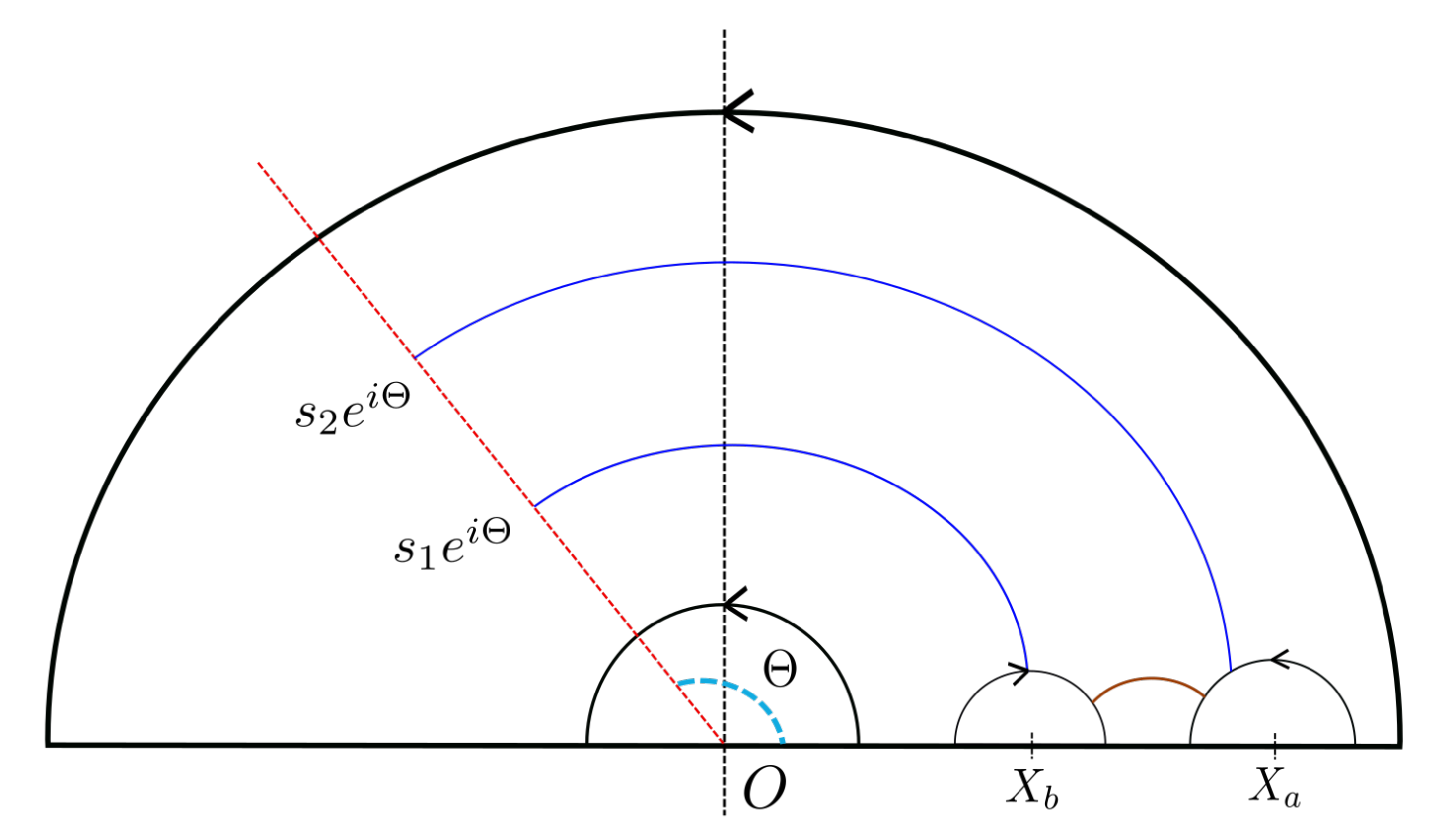}
	\caption{The covering space of the multiboundary wormhole from the inception geometry side. Considering that a candidate RT surface for radiation subregions $R_1$ and $R_2$ can pass through the EoW brane from the inception geometry to the real geometry, the blue lines are drawn, which correspond to the minimal ones anchoring on the EoW brane. It will close in the actual black hole side not shown in this picture. }
	\label{mbwent}
\end{figure}

From the lengths discussed above, it is straightforward to calculate the entanglement entropy between $R_{1}$ and the actual black hole, which is given by the minimum ``length" among all possible candidate RT surfaces as \cite{Balasubramanian:2020hfs}
\begin{align}
    S_{R_1} = \mathrm{min} \bigg[\underset{s_1,s_2}{\mathrm{min}}\bigg[ \frac{L_I (s_1, s_2)}{4 G_N'} + \frac{L_{\mathrm{bh}} (s_1, s_2)}{4 G_N} \bigg], \frac{m_1}{4 G_N'} \bigg]. \label{ee}
\end{align}
where $s_1 e^{i \Theta}$ and $s_2 e^{i \Theta}$ are the coordinates of the ``infalling geodesics" (see Fig.\ref{mbwent}). The subscript $I$ on $L_I$ indicates the length has to be computed from the \emph{inception} side. The length of the geodesic from real black hole side is given by $L_{\mathrm{bh}} (s_1, s_2)$ (as shown in Fig.\ref{wormisl}$b$, the candidate RT surface closes in the actual BH side). Note that, from either side the geodesic arcs have to meet at $[s_1,s_2]$ on the EoW brane and we have to minimise with respect to these points. This contribution dominates after the secret-sharing time whereas before that time, the length is simply given by $m_1/4G_N$ (corresponds to the dotted red geodesic between $X_a$ and $X_b$ centered semicircles in Fig.\ref{fig:mbw3}). Note the crucial difference in $G_N'$ and $G_N$. The prime indicates the inception side while the un-primed coordinates are from the real side of the black hole. Hence, initially until the secret-sharing time, the RT surface corresponding to $R_1$ stays completely in the inception side whereas after the secret-sharing time it partially includes the ``island" region.\footnote{Just as a reminder and to avoid any confusion, let us reiterate that the nomenclature reflects the fact that a part of the radiation gets access to a partial island region at this point of time and cannot get access to the full islands unless there is a quantum channel or secret-sharing between the subsystems.}
For simplicity, we choose all moduli to be of same length (as was done in \cite{Balasubramanian:2020hfs}). They are related with the inception horizon as
\begin{align}
    m_1 = m_2 = m_3 = 2 \pi r_h'\,.
\end{align}
We choose the following parametrization with $\lambda=\mu^2$, so that
\begin{align}
X_a &= \mu X_b, ~~ X_b = \frac{\sqrt{D_a D_b} (1 + \mu^2)}{\mu (\mu-1)}, \nonumber \\
D_1 &= \frac{1}{\mu}, ~~ D_2 = \mu, \nonumber \\
D_a &= 2 \bigg(\frac{\mu-1}{2 \mu} \bigg), ~~ D_b = \frac{1}{2} \bigg(\frac{\mu-1}{2 \mu} \bigg), \label{param}
\end{align} 
where $\mu$ is constant and controls the evaporation protocol. The parametrization implies
\begin{align}
    m_{1,2,3} = 2\ell' \ln \mu, ~~~~ \Rightarrow ~~~~\mu = \exp{ \left(\frac{\pi r_h'}{\ell'}\right)}\,.
\end{align}
We can immediately see that $r_h' = 0$ implies $\mu = 1$. Hence the evaporation protocol starts at $\mu = 1$, and it plays the role of ``time".
\begin{figure}
\centering
\includegraphics[scale=0.5]{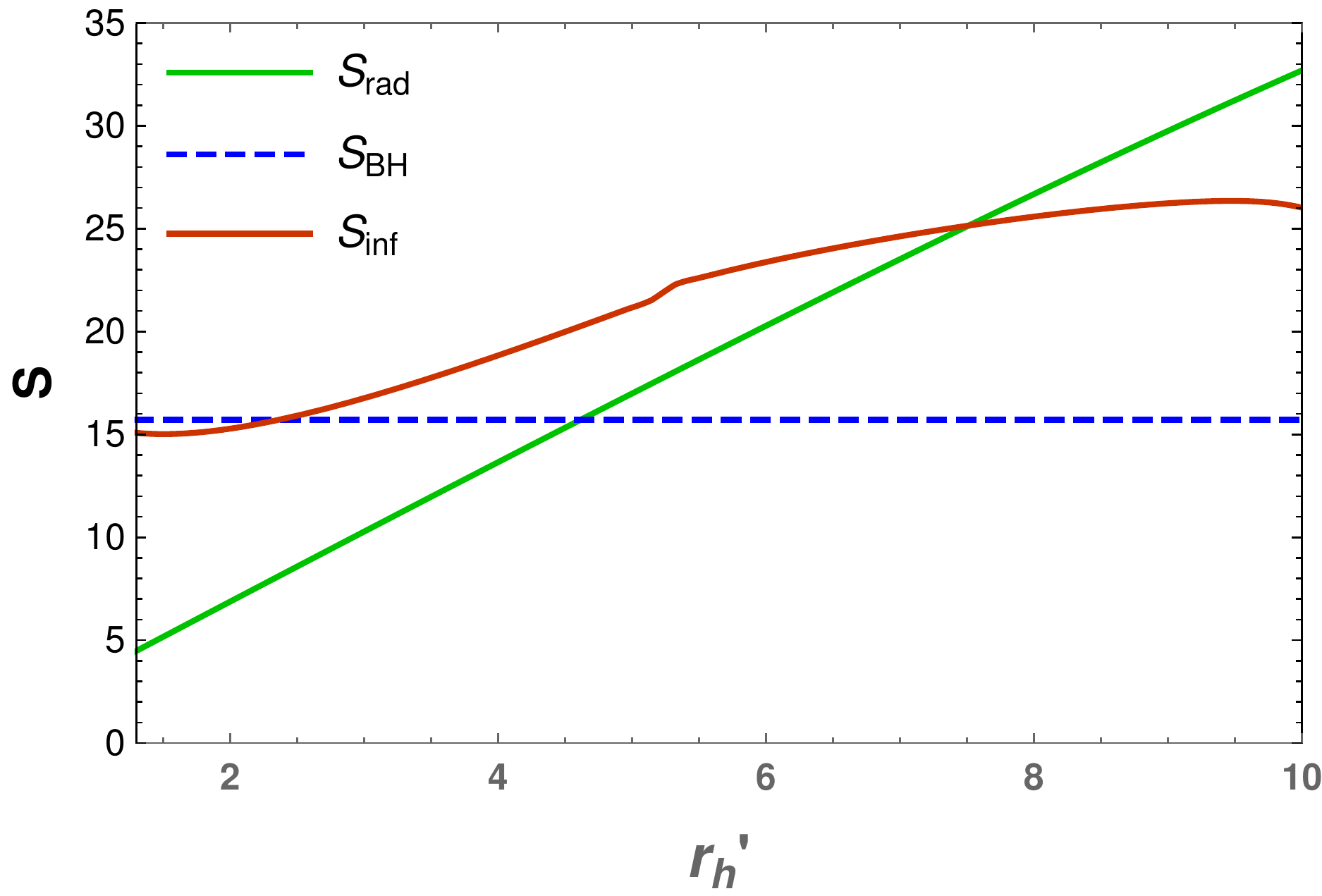} 
\caption{The Page curve according to Eqn.\eqref{ee}. We take $r_h=10$, $G_N=1$, $\ell=1$ and $\hat{c}=0.2$. The green curve indicates the entropy of the radiation subsystem $R_1$ which continues to grow. Dashed blue curve is the entropy corresponding to the real black hole horizon which stays constant. The brown curve corresponds to the infalling geodesics which goes through the EoW brane from inception side to the real side and contains the island region partially. The intersection of green curve and dashed blue curve marks the Page time whereas the the intersection of brown and green curve is termed as secret-sharing time. The colors are according to the Fig.\ref{wormhole5}.}
\label{pagecurve}
\end{figure}
Since the moduli ($m_1$, $m_2$ and $m_3$) are of the same length by construction, we can address two transitions in a single plot. If we consider the union of $R_{1}$ and $R_{2}$, the entire radiation region, there are two candidates of RT surfaces capturing the entanglement between the radiation and the black hole. We can write the island formula for this case as the following
\begin{equation}\label{fullEE}
    S(R_1 \cup R_2)=\mathrm{min} \bigg[ \frac{h}{4 G_N}, \frac{m_3}{4 G_N'} \bigg],
\end{equation}
where $m_3=2\pi r_h^{\prime}=m_1$ and $h=2\pi r_h$ similarly ($h$ being the moduli in the real-side geometry which does not change with time modelling the eternal black hole geometry in a way). Now, initially $m_3/4G_N^{\prime}$ is the RT surface for the complete radiation region. However as time passes and therefore $r_{h}^{\prime}$ increases, at the Page time, the other choice in Eq.\eqref{fullEE} becomes the minimal RT surface. This transition is shown in Fig.\ref{pagecurve} by the crossover between the green growing curve and the dotted blue constant curve. The parameter $r_h'$ plays the role of ``time" and the crossing point value (between the green and blue curves) in the $x$ axis represents the Page time. \par
Now let us consider the entanglement between one half of the radiation ($R_1$) and the black hole. It is now straightforward to compute the entanglement entropy given in Eq.\eqref{ee}. The plot is shown in Fig.\ref{pagecurve}. The crucial point here to note is that the green curve now represents $m_1/4G_{N}^{\prime}$ from Eq.\eqref{ee} due to all the three moduli in the inception side being equal by construction.\footnote{This choice gives us a way to compare the two timescales in a single plot.} The other candidate in Eq.\eqref{ee} is plotted through the red curve. The secret-sharing time is when the radiation curve (green) crosses the infalling geodesics (red). The secret-sharing time is larger than the Page time. This captures the fact that although the information of the whole island region becomes accessible to $R_{1}\cup R_2$ at Page time, the information is not uniformly distributed to all small parts of the radiation. It becomes clear from this study that one half of the radiation gets to know about the island region much after the Page time and yet only partially.
\section{Holographic subregion complexity}\label{sec3}
In this section, we compute the subregion complexity for the evolving mixed state dual to the entanglement wedge for half of the radiation system. Note that the entanglement wedge is a co-dimension one region similar to the volume below the RT surface. Since the entanglement wedge is dual to the density matrix of a mixed state, the volume of the region has also been considered as the complexity of a mixed state extending the ``complexity$=$ volume" \cite{Susskind:2014moa, Susskind:2014rva} for pure states. The volume dual to a subregion is usually dubbed as the ``subregion complexity" \cite{Agon:2018zso, Alishahiha:2018lfv,Chen:2018mcc}, which has been investigated for diverse scenarios \cite{Alishahiha:2015rta, Abt:2017pmf, Abt:2018ywl, Bhattacharya:2019zkb,Auzzi:2021nrj, Baiguera:2021cba,Sato:2021ftf, Bernamonti:2021jyu}. In holography, the subregion complexity of a boundary region $A$ with a bulk minimal surface $\gamma_A$ is denoted as
\begin{equation}
    C_V=\frac{\mathcal{V}\left(\gamma_A\right)}{8\pi \ell G_N}.
\end{equation}
where $\mathcal{V} (\gamma_A)$ is the enclosed volume by the boundary subregion $A$ and the minimal surface $\gamma_A$, and $\ell$ is the AdS radius. In the model that we are studying, we will compute the volumes between $R_1$ and its corresponding RT surface at different times. Since the multiboundary wormhole models are well understood as timeslices and for AdS$_3$ co-dimension one regions are two dimensional, the volumes are actually be the areas in the pictures. This study is in a similar spirit to that of other recent studies of the subregion complexity for the doubly holographic models of islands both for evaporating and eternal black holes \cite{Bhattacharya:2020uun, Hernandez:2020nem, Bhattacharya:2021jrn, Sato:2021ftf}.

\subsection{Complexity before secret-sharing time}

\begin{figure}
\centering
\includegraphics[scale=0.52]{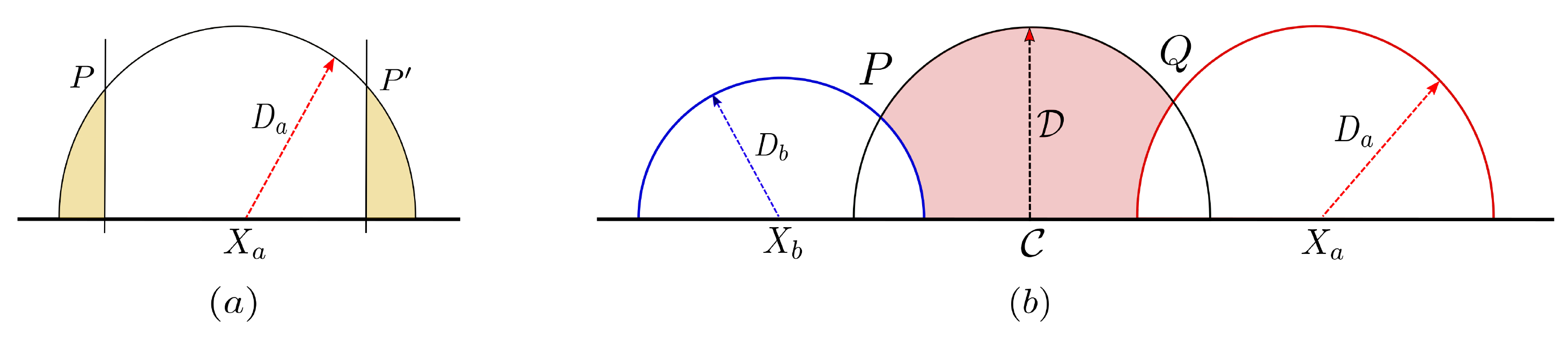} 
\caption{$(a)$ Volume enclosed by a vertical line at $P$ or $P'$ and a half circle. $(b)$ Volume under $m_1$.}
\label{fig5}
\end{figure}

Here we compute the volume under the relevant geodesic before secret sharing time, which is the volume shown in the shaded region in Fig.\ref{fig5}$b$. Typically in AdS$_3$, if the bulk curves are anchored to the boundary then the volumes are given by the boundary subregion's size $x$ divided by some UV cutoff $\epsilon$
\begin{equation}
    \mathcal{V}=\frac{x}{\epsilon}+\alpha
\end{equation}
where $\alpha$ are angles that depend upon the topology (Euler characteristics) of the region whose volumes we try to compute \cite{Abt:2017pmf, Bhattacharya:2020uun}. However, in the model under study, the situation needs to be tackled a bit more carefully since $m_1$ does not reach the boundary and ends on the body of the wormhole. In what follows, we discuss the detailed computations of how such volumes are computed with relevant figures.

The UHP metric is given by (we will momentarily set $\ell'=1$)
\begin{equation}
ds^2=\frac{dx^2+dy^2}{y^2}\,.
\end{equation}
We compute the volume under a semicircle of radius  $R$ which is anchored at the boundary and is given by
\begin{align}
    \mathcal{V}=\int\int\frac{dx dy}{y^2}= \frac{2R}{\epsilon} -\pi\,,
\end{align}
where $\epsilon$ is the location of the cutoff surface, which ensures that we get a finite answer for the volume.

It will be useful for us to compute the volume between a half circle and a vertical line that cut the center at $x>0$ with respect to the centre of the circle. This volume is depicted in the Fig.\ref{fig5}$a$ and given by
\begin{eqnarray}
\mathcal{V}(x)&=& \frac{D_a-x}{\epsilon}-\frac{\pi}{2}+\sin^{-1}\left(\frac{x}{D_a}\right)\,.
\label{3.4}
\end{eqnarray}
Using $\eqref{3.4}$ we find the volume under $m_1$
\begin{align}
\begin{split}
\mathcal{V}_{t<t_s}=\pi&-\frac{D_b+D_a-(X_a-X_b)}{\epsilon}-\sin^{-1}\left(\frac{P_x-X_b}{D_b}\right)-\sin^{-1}\left(\frac{\mathcal{C}-P_x}{\mathcal{D}}\right)\\
&-\sin^{-1}\left(\frac{X_a-Q_x}{D_a}\right)-\sin^{-1}\left(\frac{Q_x-\mathcal{C}}{\mathcal{D}}\right)\,.
\end{split}
\end{align}
$\mathcal{C}$ is the center and $\mathcal{D}$ is the radius of the moduli $m_1$
\begin{equation}
\mathcal{C}=\frac{X_a+X_b}{2},\,\,\,\,\,\,\,\,\,\,\,\,\,
\mathcal{D}=\frac{1}{2}\sqrt{(X_a-X_b)^2-4D_a D_b}\,.
\end{equation}
where $P_x$, $P_y$, $Q_x$, and $Q_y$ are given by the relation $\eqref{2.13}$ and $\eqref{2.14}$.
After further simplification we find that the other angular parts exactly cancel $\pi$ and get the volume of the shaded region in Fig.\ref{fig5}$b$ as,
\begin{eqnarray}
\mathcal{V}_{t<t_s}=\frac{X_a-X_b-(D_b+D_a)}{\epsilon}\,,
\end{eqnarray}
As before, we take the symmetric setup with $m_1 = m_2 = m_3$. Choosing the parametrization \eqref{param} we compute the volume
\begin{align}
\mathcal{V}_{t<t_s} = \frac{(\mu-1)(\mu-2)(2 \mu -1)}{4 \mu^2 \epsilon},
\end{align}
where $\mu = \exp(\pi r_h'/\ell')$ and $r_h'$ being the radius of the throat horizons of $R_1$ (or $R_2$ or $R_3$) and acts as a proxy of time. Hence, the subregion complexity is given by
\begin{align}
C_{t<t_{s}} = \frac{\ell'\mathcal{V}_{t<t_s}}{8 \pi G_N'} = \frac{\ell'(\mu-1)(\mu-2)(2 \mu -1)}{32 \mu^2 \epsilon \pi G_N'}\,.
\end{align}
Here $\ell'$ and $G_N'$ are the AdS radius and Newton's constant for the inception geometry respectively. The plot of complexity with respect to `time' (kept track of in terms of $r_{h}^{\prime}$) is shown in Fig.\ref{fig:volgrow}. Note that the volume is ever-growing, supporting the notion that although entanglement does not grow forever, complexity does. This is usually noted as a consequence of the Hawking's original calculation being applicable to complexity of the radiation state.
\begin{figure}[t]
\centering
\includegraphics[scale=0.52]{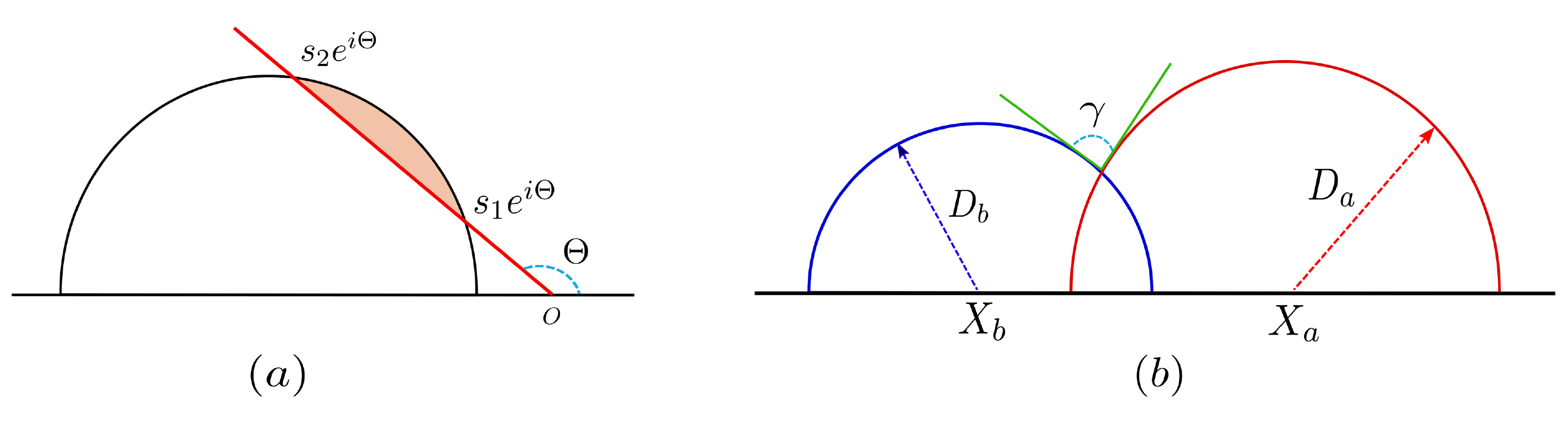} 
\caption{(a) Area enclosed by the brane and a part of infalling geodesic. (b) Angle between two intersecting geodesics.}
\label{infall}
\end{figure}
\subsection{Complexity after secret-sharing time}
\begin{figure}[t]
	\centering
	\includegraphics[scale=0.3]{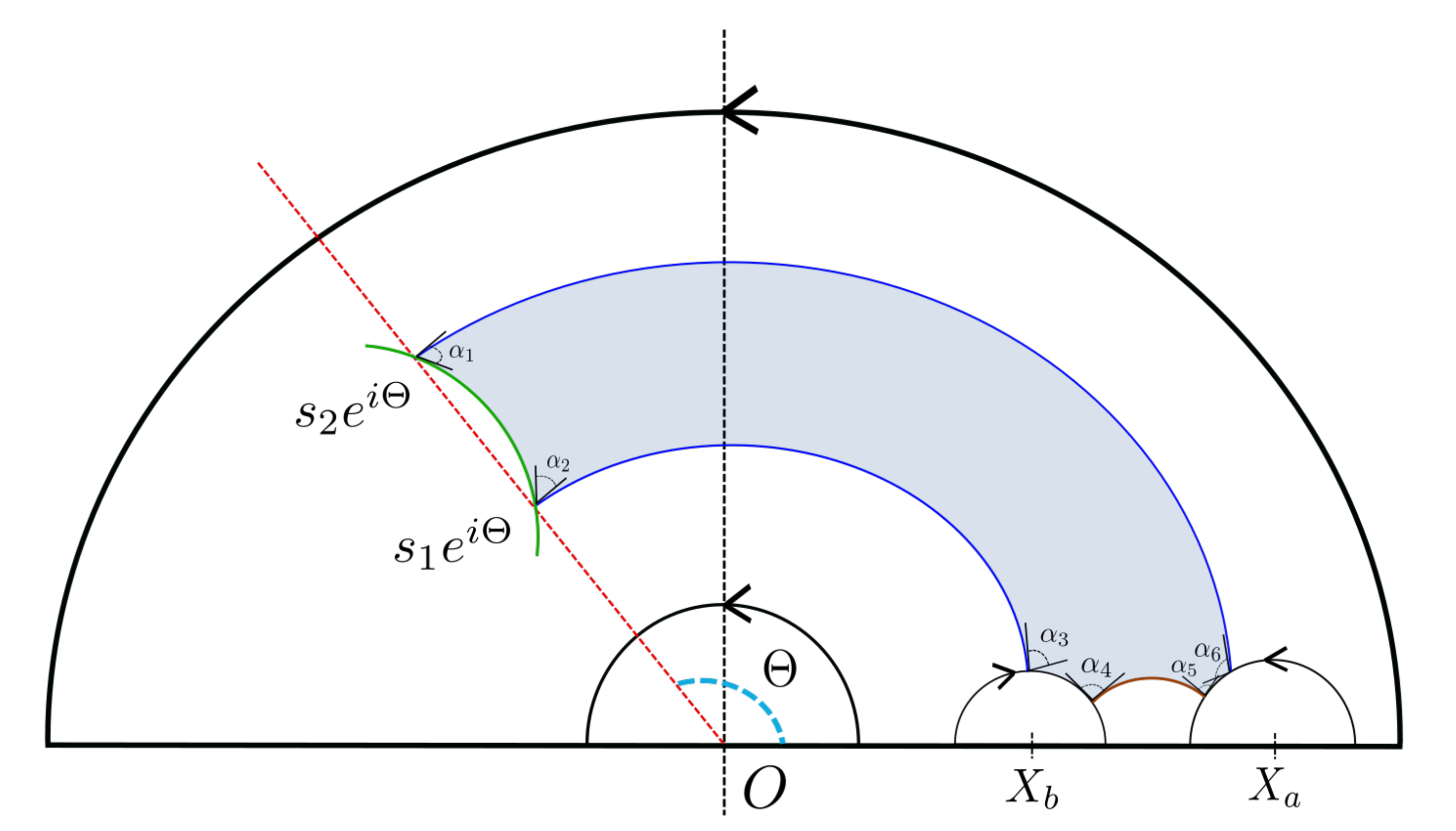}
	\caption{The blue shaded region corresponds to the volume between a set of geodesics in the inception sides that is helpful for calculations. To compute the full volume in the inception side, one needs to add the volume between $s_1$ and $s_2$ on the EoW brane and the green curve.}
	\label{mbw}
\end{figure}

\begin{figure}
		\centering
		\begin{subfigure}[b]{0.50\textwidth}
			\centering
			\includegraphics[width=\textwidth]{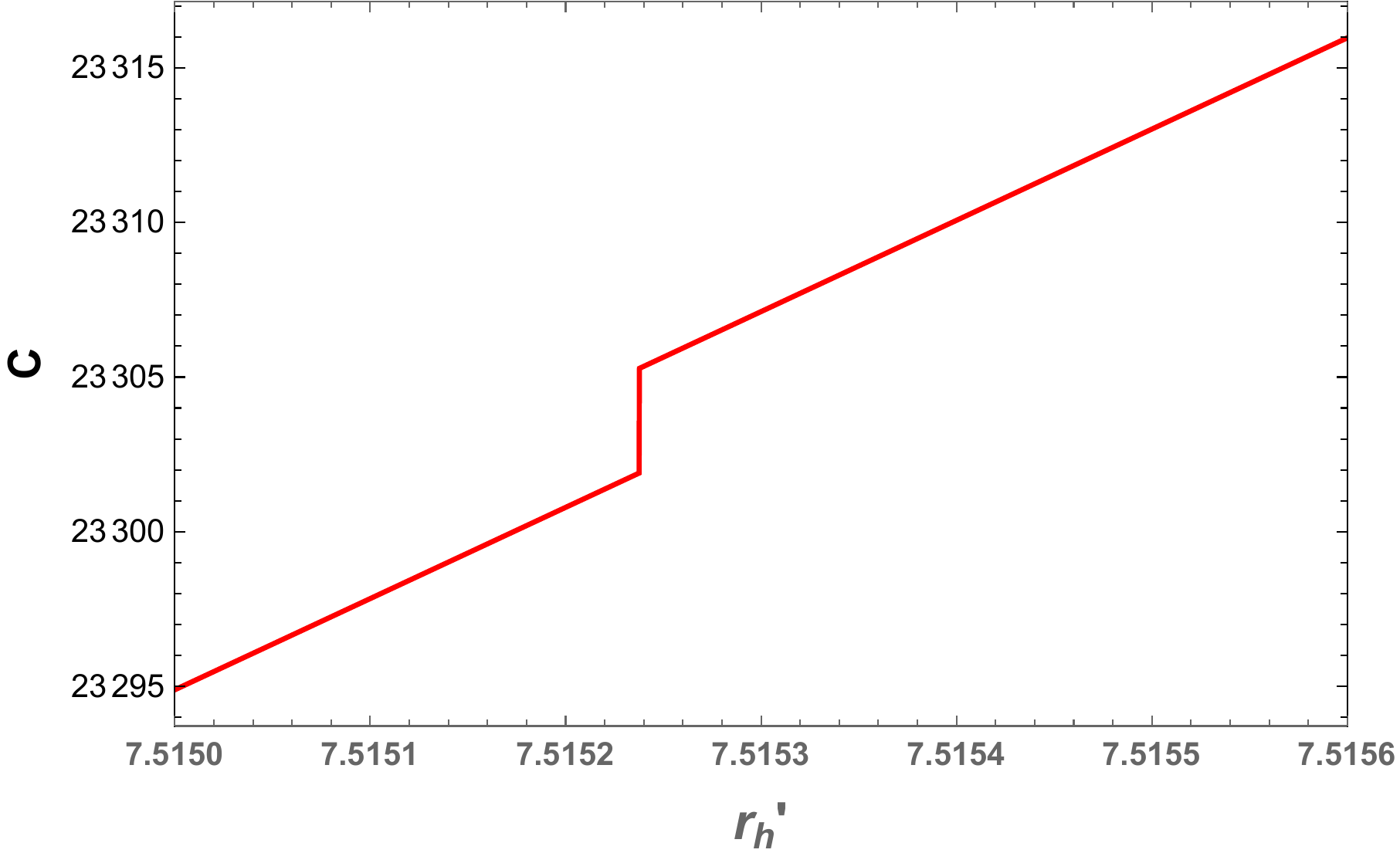}
			\caption{}
			\label{fig:volgrow}
		\end{subfigure}
		\hfill
		\begin{subfigure}[b]{0.47\textwidth}
			\centering
			\includegraphics[width=\textwidth]{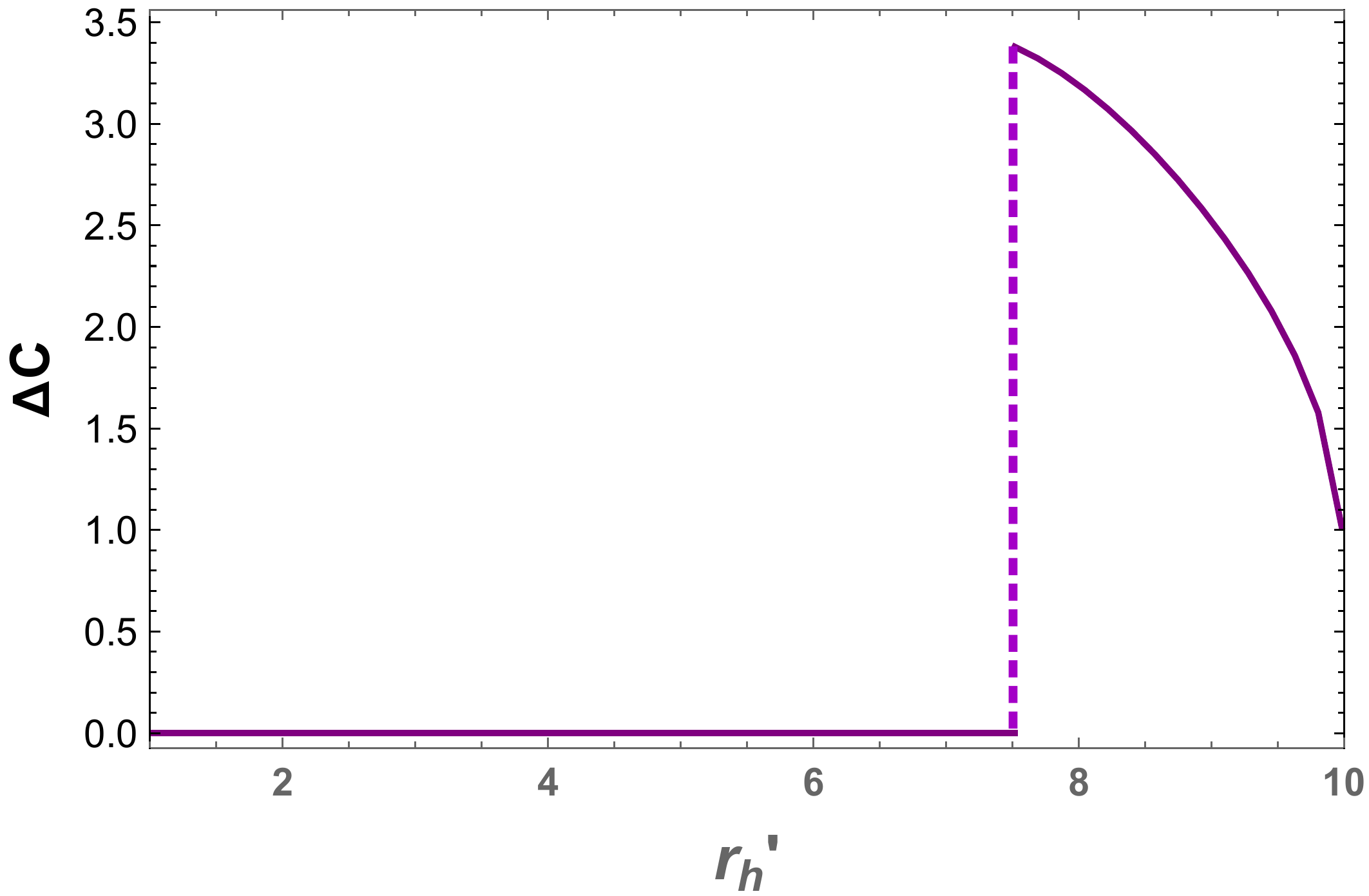}
			\caption{}
			\label{fig:volchange}
		\end{subfigure}
		\caption{Variation of subregion complexity with respect to the throat horizon of radiation subsystem $R_1$. Complexity shows a discontinuous jump at secret-sharing time. (b) Starting from the secret-sharing time the complexity of the partial island $\Delta C$ decreases gradually.}
		\label{fig:volvol}
\end{figure}

We now compute complexity after secret sharing time $t_s$. Contribution to the Complexity comes from real side as well as the inception side. For computational simplicity we divide the inception volume into three parts:- $i)$ Volume under $m_1$, $ii)$ Volume enclosed by $m_1$ and the geodesic passing through $s_1e^{i\Theta}$ and $s_2e^{i\Theta}$ shown in Fig.\ref{mbw}, $iii)$ Volume enclosed by the EoW brane and the geodesic passing through $s_1e^{i\Theta}$ and $s_2e^{i\Theta}$ shown in Fig.\ref{infall}$a$. We have already discussed the volume under $m_1$
\begin{align}
\mathcal{V}_{t<t_s} = \frac{(\mu-1)(\mu-2)(2 \mu -1)}{4 \mu^2 \epsilon}\,.
\end{align}
The volume of the shaded region in Fig.\ref{infall}$a$,
\begin{equation}
\mathcal{V}_I=-\sin^{-1}\left(\frac{s_1\cos\Theta -X}{D}\right)+\sin^{-1}\left(\frac{s_2\cos\Theta -X}{D}\right)-\cot\Theta\log\left(\frac{s_2}{s_1}\right),
\end{equation}
where
\begin{eqnarray}
X&=&-\frac{1}{2}(s_1+s_2)\sec{\Lambda}\,,\\
D&=&\frac{1}{2}\sqrt{s_1^2+s_2^2-2 s_1 s_2 \cos(2\Lambda)}\sec\Lambda\,,\\
\Theta&=&\pi-\Lambda\,.
\end{eqnarray}
Now, we use Gauss-Bonnet formula to compute the volume of the shaded region in Fig.\ref{mbw},
\begin{equation}
\tilde{\mathcal{V}}=4\pi-\sum_{i=1}^{6}\alpha_i\,,
\end{equation}
where $\alpha$'s are the interior angles of the hyperbolic polygon depicted in Fig.\ref{mbw}.
In order to calculate $\alpha$'s we need to compute the angle between two intersecting semicircles.  This angle is depicted in Fig.\ref{infall}$b$. We take two circles with centers at $X_b$ and $X_a$ and with radius $D_b$ and $D_a$ respectively. The angle between these two circles is,
\begin{equation}
\gamma=\pi-\left(\sin^{-1}\left(\frac{(X_a-X_b)^2+D_b^2-D_a^2}{2D_b(X_a-X_b)}\right)+\sin^{-1}\left(\frac{(X_a-X_b)^2-D_b^2+D_a^2}{2D_a(X_a-X_b)}\right)\right)\,.
\label{3.17}
\end{equation}
Using $\eqref{3.17}$ we find $\alpha$'s,
\begin{align}
\begin{split}
&\alpha_1 = \sin^{-1}\left(\frac{(X_J-X)^2+D^2-D_J^2}{2D(X_J-X)}\right)+\sin^{-1}\left(\frac{(X_J-X)^2-D^2+D_J^2}{2D_J(X_J-X)}\right),\\&
\alpha_2= \pi-\left(\sin^{-1}\left(\frac{(X_I-X)^2+D^2-D_I^2}{2D(X_I-X)}\right)+\sin^{-1}\left(\frac{(X_I-X)^2-D^2+D_I^2}{2D_I(X_I-X)}\right)\right),\\&
\alpha_3 =\pi-\left(\sin^{-1}\left(\frac{(X_b-X_I)^2+D_I^2-D_b^2}{2D_I(X_b-X_I)}\right)+\sin^{-1}\left(\frac{(X_b-X_I)^2-D_I^2+D_b^2}{2D_b(X_b-X_I)}\right)\right),\\&
\alpha_6 =\sin^{-1}\left(\frac{(X_a-X_J)^2+D_J^2-D_a^2}{2D_J(X_a-X_J)}\right)+\sin^{-1}\left(\frac{(X_a-X_J)^2-D_J^2+D_a^2}{2D_a(X_a-X_J)}\right)
\end{split}
\end{align}
where $X_I$, $X_J$, $D_I$ and $D_J$ are given by
\begin{equation}  
X_I=\frac{2 s_2 \cos\Theta \left(s_1^2 X_a-X_b \left(D_a D_b+X_a X_b\right)\right)+2 D_a X_a D_b X_b+D_a^2
   D_b^2-\left(X_a^2+s_2^2\right) \left(s_1-X_b\right) \left(X_b+s_1\right)}{2 \left(-\cos \Theta
   \left(s_2 D_a D_b+2 s_2 X_a X_b+s_1 \left(X_a^2+s_2^2\right)\right)+D_a X_a D_b+X_b
   \left(X_a^2+s_2^2\right)+s_1 s_2 X_a \cos 2 \Theta +s_1 s_2 X_a\right)}\,,\nonumber
   \end{equation}
   \begin{equation}
    X_J=\frac{2 s_1 \cos\Theta  \left(s_2^2 X_b-X_a \left(D_a D_b+X_a X_b\right)\right)+2 D_a X_a D_b X_b+D_a^2
   D_b^2-\left(s_2-X_a\right) \left(X_a+s_2\right) \left(X_b^2+s_1^2\right)}{2 \left(-\cos \Theta
   \left(s_1 D_a D_b+2 s_1 X_a X_b+s_2 \left(X_b^2+s_1^2\right)\right)+D_a D_b X_b+X_a
   \left(X_b^2+s_1^2\right)+s_1 s_2 X_b \cos 2 \Theta+s_1 s_2 X_b\right)}\,,\nonumber
   \end{equation}
   \begin{equation}
D_I=\sqrt{\left(X_I-s_1 \cos \Theta \right)^2+s_1^2 \sin ^2 \Theta},\,\,\,\, D_J=\sqrt{\left(X_J-s_2 \cos \Theta \right)^2+s_2^2 \sin ^2 \Theta}\,.
\end{equation}
Also, we find $\alpha_4+\alpha_5=\pi$. Finally, we compute the volume $\mathcal{V}_R$, which is the volume enclosed by the brane and a part of infalling geodesic which has the endpoints on the brane in the real side. This volume is same as the volume $\mathcal{V}_I$ enclosed by EoW brane and the geodesic passing through $s_1e^{i\Theta}$ and $s_2e^{i\Theta}$. After collecting all the terms, we find complexity after secret sharing time is,
\begin{equation}
C_{t>t_{s}}=C_{t<t_s}+\frac{\ell'(\tilde{\mathcal{V}}+\mathcal{V}_I)}{8\pi G_N'}+\frac{\ell \,\mathcal{V}_R}{8\pi G_N}\,.
\end{equation}
$C_{t<t_s}$ is a divergent quantity as $\epsilon\rightarrow0$ so we define a finite quantity $\Delta C$ as follows,
\begin{equation}
\Delta C=
    \begin{cases}
    0&\,\,\,t\leq t_s\\
    \frac{\ell'(\tilde{\mathcal{V}}+\mathcal{V}_I)}{8\pi G_N'}+\frac{\ell\,\mathcal{V}_R}{8\pi G_N}&\,\,\,t>t_s
    \end{cases}
\end{equation}
The complexity of the partial island is shown in Fig.\ref{fig:volchange}.

\section{Classical Markov channel and geometric secret-sharing:}\label{Markovapp}

Let us consider a pure tripartite state whose density matrix is written as $\rho_{ABC}$. In general, if we trace out the degrees of freedom corresponding to region $C$, we would be left with the reduced density matrix of the mixed state  $\rho_{AB}$. Now, let us assume a map $\mathcal{R}_{B\rightarrow BC}$ which, as the notation suggests, maps the system $B$ into $BC$ (note that this does not guarantee that the full region $BC$ is exhausted through this map). We can think of this map as some quantum channel between $B$ and $C$. If an observer on $B$ has access to $\rho_{AB}$ and acts with this channel ($\mathcal{R}_{B\rightarrow BC}$) on the reduced mixed density matrix $\rho_{AB}$, a tripartite state $\Tilde{\rho}_{ABC}$ could be produced which will have support over the whole system. This channel is typically known as the \emph{Markov recovery channel} \cite{Hayden:2021gno}
\begin{equation}
    \Tilde{\rho}_{ABC}= \mathcal{R}_{B\rightarrow BC} \,(\rho_{AB}).
\end{equation}
The quantum distance defined through the fidelity between the density matrices of the total system $\rho_{ABC}$ and the newly produced state $\Tilde{\rho}_{ABC}$ measures how well the recovery channel can recover the original state. The fidelity between two density matrices $\rho$ and $\sigma$ is defined as 
\begin{equation}
    \mathcal{F}(\rho,\sigma)= \mathrm{Tr}\left( \sqrt{\sqrt{\rho}\sigma\sqrt{\rho}}\right)^2, ~~~ \text{satisfying} ~~~ 0 \leq \mathcal{F}(\rho,\sigma)\leq 1.
\end{equation}
This measures the overlap between two density matrices in consideration. Therefore, the more the fidelity value (closer to unity), the closer two density matrices are to each other.

For a good quantum recovery channel, the expectation is that there exists a Markov chain ($A\rightarrow B \rightarrow C$), which exactly reproduces the original density matrix. For more details on these, we refer \cite{Hayden:2021gno} which has addressed recovery in the context of the reflected entropy and the mutual information in holography. The fidelity between the exact state and the recovery state lower bounds the mutual information which in turn lower bounds the reflected entropy. The difference between reflected entropy and the mutual information is always positive and termed as \emph{Markov gap}. In the following, we try to argue how a similar concept like \emph{Markov gap} can be associated with our study in the geometric secret-sharing model.
\begin{figure}[t]
	\centering
	\includegraphics[scale=0.48]{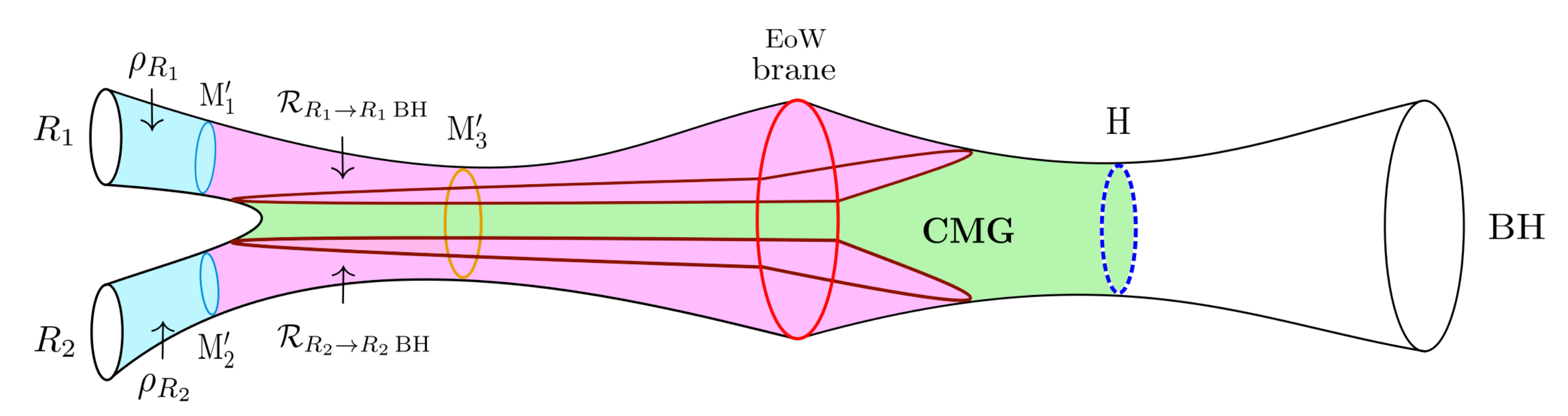}
	\caption{Classical Markov recovery. The pink region denotes the volume enclosed by the infalling geodesics from radiation subsystem $R_1$ and $R_2$.
	The shaded green region is the classical Markov gap explained in the main text. Here we assume that two observers sitting at $R_1$ and $R_2$ exits can communicate classically, but neither they have the full control over both the exists simultaneously.}
	\label{markov}
\end{figure}
In our case, we do not talk about Markov gaps from the perspective of the reflected entropy. However, it is worth mentioning that such a study can be possible since reflected entropy in this particular model has been studied previously in \cite{Li:2020ceg}. In our study, we restrict ourselves to the standard definition of the Markov gap mentioned above. We see that our radiation subsystems $R_{1,2}$ are very similar to the $A$ and $B$ subsystems mentioned above, whereas the original black hole exit plays the role of subsystem $C$.

The statement of a quantum Markov recovery channel is  to act on $\rho_{AB}$ with a recovery channel $\mathcal{R}_{B\rightarrow BC}$. However, from an observer situated at the exit $R_1$, it is also nontrivial to operate on the full $\rho_{R_1\cup R_2}$ (equivalently the EW of $R_1\cup R_2$) unless there already exists a quantum channel between $R_1$ and $R_2$ through which it can access the information of the region EW$(R_1\cup R_2)-$EW$(R_1)-$EW$(R_2)$. We, on the other hand, do not assume existence of such a channel. Hence, in our case we assume the recovery map $\mathcal{R}_{R_1\rightarrow R_1 \, \mathrm{BH}}$ to be acting on $\rho_{R_1} \otimes \rho_{R_2}$ and similarly from the $R_{2}$ side, another recovery map $\mathcal{R}_{R_2 \rightarrow R_2 \, \mathrm{BH}}$ to be acting on $\rho_{R_1}\otimes \rho_{R_2}$. Here the subscript ``BH" stands for black hole. We call this as \emph{classical Markov recovery channel}
\begin{align}
    \rho_{R_1 \,R_2 \, \mathrm{BH}}^{\prime}= \mathcal{R}_{R_1 \rightarrow R_1 \, \mathrm{BH}} (\rho_{R_1}\otimes \rho_{R_2}) \otimes \mathcal{R}_{R_2 \rightarrow R_2 \, \mathrm{BH}} (\rho_{R_1}\otimes \rho_{R_2}).
\end{align}
The role of the channels $\mathcal{R}_{R_1 \rightarrow R_1 \, \mathrm{BH}}$ and $\mathcal{R}_{R_2 \rightarrow R_2 \, \mathrm{BH}}$ are played by the QES at various times. For example, if we fix the observer to $R_1$ and try to look beyond its trivial EW by asking something extra. The added non-trivial extension of its possible RT surfaces to cross the EoW brane, which was the main point of \cite{Balasubramanian:2020hfs}, is also what plays the crucial role in our definition of the classical recovery channels. Unless this extension was there EW(${R_1}$) would never go beyond the usual RT surface. A symmetrically placed channel can also be formed from the perspective of the $R_2$ side. 

Finally we can define the \emph{classical Markov gap (CMG)} as $\mathcal{F}(\rho_{R_1 \,R_2 \, \mathrm{BH}},\rho^{\prime}_{R_1 \,R_2 \, \mathrm{BH}})$, the fidelity between the original density matrix and the density matrix produced by a classical Markov recovery channel. Since the entanglement wedges (codimension $1$ regions, similar to the volumes) are usually thought to carry information equivalent to the corresponding density matrices, the volume of the 
\begin{equation*}
    \text{\emph{full island}}- 2 \,\times\text{\emph{partial islands}}
\end{equation*}
 region measures the size of the fully quantum secret sharing channel. The classical Markov recovery channels can not access any of this region because these channels are completely classical. A perfect quantum Markov chain ($\tilde{\rho}_{R_1 \,R_2 \, \mathrm{BH}}=\rho_{R_1 \,R_2 \, \mathrm{BH}}$) can be made possible only if one can have access to a fully quantum secret-sharing channel.

\section{Discussions:} \label{secdiscuss}

In this paper, we study the geometric secret sharing model where part of the radiation can at most access only a part of the island region and only after the secret-sharing time, which is greater than the Page time. Therefore, the full quantum secret of the island region is only shared with the full radiation system. The region which can never be accessed by individual observers communicating through classical channels should be understood as a quantum phenomenon in this model, which makes this region extremely special.

We elucidate the above scenario by applying machinery from multiboundary wormholes in AdS$_3$. The areas provide the expected behaviour, as shown in \cite{Balasubramanian:2020hfs}. The RT surfaces for half of the radiation goes through a transition at the so-called \textit{secret-sharing time}. We, therefore, extend the area study to the corresponding volumes, in order to study the corresponding subregion complexities, where the topologies of the regions under study appear in the computations. Another important change due to the passing of RT surfaces through the EoW brane is that one has to be careful in assigning different values to the Newton's constant and AdS length scale to the inception- and the real-side geometries. Our conclusions are listed below.

\begin{enumerate}
	\item The volume accessible to the complete radiation subsystem keeps increasing with time. This volume is the UV cut-off-dependent divergent volume. However, after the Page time, a new volume gets added to the radiation subsystem, which is finite. We call this volume the \textit{island volume}. As argued in our previous work \cite{Bhattacharya:2021jrn}, this volume grows in time and represents the complexity of auto-purification\footnote{Here, auto-purification means that one would not need to add any ancilla system to purify for these parts of the mixed state under study. Due to both partner modes (Hawking quanta inside and outside black hole), these modes form a pure-state (one could think of it forming an eigenstate present in the classical probabilistic mixture forming the mixed state) within the mixed state. In the secret-sharing model, it is easy to point out. The island is part of the real geometry on the right-hand side of the EoW brane. However, since the model deals with a purification done by the inception geometry, the new volume also contains a region in the inception geometry, representing the modes that purify the island modes.} of certain modes that are purified after the island is encoded in the entanglement wedge of the radiation subsystem.
	
	\item When dealing with one half of the radiation subsystem, we again find a similar behaviour of the volumes. There is one highly divergent piece due to the UV cut-off dependence in the multiboundary of AdS$_{3}$ boundaries. However, after the secret-sharing time, a new volume gets added to the entanglement wedge of the $R_1$ subregion. This newly added volume again consists of parts  both from the real- and the inception-side geometries. Hence, this volume again represents some purification taking place and resulting in the jump. However, when we study the time dependence of this \textit{partial island volume} after secret-sharing time, we find that it typically decreases with time. 
	
	\item One should be careful and understand that this volume change is much less than the overall divergent volume, which keeps growing in time as the size of the throat horizon of $R_1$ keeps increasing. However, due to this decreasing nature of \textit{partial island volume}, the slope of the overall volume is expected to decrease after the secret-sharing time (whereas it increases if one considers the total radiation subsystem $R=R_1\cup R_2$).
	
	\item $R_1$ and $R_2$ are considered to be of equal size in the model, and therefore, both would get access to symmetrically placed \textit{partial island} regions after secret-sharing time. However, the union of the \textit{partial islands} does not form the complete \textit{island} region. This region is particularly interesting since these are the modes that are not accessible through a classical channel. Observers $\mathcal{O}_{R_1}$ and $\mathcal{O}_{R_2}$, living on $R_1$ and $R_2$, respectively, need to share some kind of quantum secret between themselves to access over this region. We refer to this volume as the ``\textit{secret-sharing volume}".

	\item The volumes and the change of subregion complexity have been previously argued to be dual to the fidelity of two perturbed quantum states \cite{Miyaji:2015woj,Alishahiha:2017cuk,Flory:2017ftd, Bhattacharya:2019zkb, Banerjee:2017qti}. In the model we study, the volume of the region \emph{islands}$- 2 \,\times$\emph{partial islands} can similarly be understood as the difference between two states where one of them can only be constructed through a quantum (secret-sharing) channel, and the other is the contribution of the fully classical channel. The difference between these kind of states is typically expressed through the Markov gaps in quantum information theory. It measures the fidelity between the density matrices of two states constructed through quantum and classical channels, respectively. A version of these Markov gaps has recently been addressed in holographic systems in \cite{Hayden:2021gno}. However, we use a little different definition (which we call the ``classical" Markov gap) than theirs, as explained in more details section \ref{Markovapp}.  
\end{enumerate}
Evolving volumes and, therefore, the subregion complexities in this picture can indeed capture the \textit{secret-sharing} transition of entanglement entropies. In addition, by looking at the nature of the complexity plot, we figure out a way to distinguish whether we are computing the complexity of the total radiation or a part of it. In the case of total radiation, the jump of complexity at Page time grows in time, whereas for half of the radiation, the complexity jump takes place at a secret-sharing time and decreases over time. This difference can also be understood as a distinguishing factor between the Page transition and the secret-sharing transition.

It would be worthwhile to understand the secret-sharing in more diverse situations e.g., the braneworld models \cite{Hernandez:2020nem, Geng:2020fxl} and the moving mirror models \cite{Akal:2020twv}. Another possible direction is to use the duality between density matrix and entanglement wedge (co-dimension one volume) to recast the Markov recovery maps and quantify the Markov gaps in various possible models (e.g., \cite{Chu:2021gdb, Li:2020ceg, Chandrasekaran:2020qtn}) more concretely. In these models, both classical and quantum Markov gaps can be accessed using the ideas introduced in \cite{Hayden:2021gno}. As the jump of complexity happens to be a universal phenomenon due to the change of preferred RT at a certain time (both for Page curve and secret sharing curve), it would be interesting to understand this jump from the perspective of holographic bit thread \cite{Freedman:2016zud, Pedraza:2021mkh, Pedraza:2021fgp}. 

\section*{Acknowledgements}
 The authors would like to thank Arjun Kar for useful discussions. A.B.$(1)$ is supported by the Institute of Eminence endowed IISc postdoctoral fellowship. A.B.$(2)$ is supported by  Start Up Research Grant (SRG/2020/001380) by Department of Science \& Technology Science and Engineering Research Board (India). P.N. acknowledges University Grants Commission (UGC), Government of India for providing financial support. A.K.P. is supported by the Council of Scientific \& Industrial Research (CSIR) Fellowship No. $09/489(0108)/2017$-EMR-I.

\bibliographystyle{JHEP}

\begin{thebibliography}{10}
	
	\bibitem{Page:1993wv}
	D.~N. Page, \emph{{Information in black hole radiation}},
	\href{http://dx.doi.org/10.1103/PhysRevLett.71.3743}{\emph{Phys. Rev. Lett.}
		{\bf 71} (1993) 3743--3746},
	[\href{https://arxiv.org/abs/hep-th/9306083}{{\tt hep-th/9306083}}].
	
	\bibitem{Page:2013dx}
	D.~N. Page, \emph{{Time Dependence of Hawking Radiation Entropy}},
	\href{http://dx.doi.org/10.1088/1475-7516/2013/09/028}{\emph{JCAP} {\bf 1309}
		(2013) 028}, [\href{https://arxiv.org/abs/1301.4995}{{\tt 1301.4995}}].
	
	\bibitem{Engelhardt:2014gca}
	N.~Engelhardt and A.~C. Wall, \emph{{Quantum Extremal Surfaces: Holographic
			Entanglement Entropy beyond the Classical Regime}},
	\href{http://dx.doi.org/10.1007/JHEP01(2015)073}{\emph{JHEP} {\bf 01} (2015)
		073}, [\href{https://arxiv.org/abs/1408.3203}{{\tt 1408.3203}}].
	
	\bibitem{Penington:2019npb}
	G.~Penington, \emph{{Entanglement Wedge Reconstruction and the Information
			Paradox}}, \href{http://dx.doi.org/10.1007/JHEP09(2020)002}{\emph{JHEP} {\bf
			09} (2020) 002}, [\href{https://arxiv.org/abs/1905.08255}{{\tt 1905.08255}}].
	
	\bibitem{Almheiri:2019hni}
	A.~Almheiri, R.~Mahajan, J.~Maldacena and Y.~Zhao, \emph{{The Page curve of
			Hawking radiation from semiclassical geometry}},
	\href{http://dx.doi.org/10.1007/JHEP03(2020)149}{\emph{JHEP} {\bf 03} (2020)
		149}, [\href{https://arxiv.org/abs/1908.10996}{{\tt 1908.10996}}].
	
	\bibitem{Almheiri:2019psf}
	A.~Almheiri, N.~Engelhardt, D.~Marolf and H.~Maxfield, \emph{{The entropy of
			bulk quantum fields and the entanglement wedge of an evaporating black
			hole}}, \href{http://dx.doi.org/10.1007/JHEP12(2019)063}{\emph{JHEP} {\bf 12}
		(2019) 063}, [\href{https://arxiv.org/abs/1905.08762}{{\tt 1905.08762}}].
	
	\bibitem{Ryu:2006bv}
	S.~Ryu and T.~Takayanagi, \emph{{Holographic derivation of entanglement entropy
			from AdS/CFT}},
	\href{http://dx.doi.org/10.1103/PhysRevLett.96.181602}{\emph{Phys. Rev.
			Lett.} {\bf 96} (2006) 181602},
	[\href{https://arxiv.org/abs/hep-th/0603001}{{\tt hep-th/0603001}}].
	
	\bibitem{Ryu:2006ef}
	S.~Ryu and T.~Takayanagi, \emph{{Aspects of Holographic Entanglement Entropy}},
	\href{http://dx.doi.org/10.1088/1126-6708/2006/08/045}{\emph{JHEP} {\bf 08}
		(2006) 045}, [\href{https://arxiv.org/abs/hep-th/0605073}{{\tt
			hep-th/0605073}}].
	
	\bibitem{Hubeny:2007xt}
	V.~E. Hubeny, M.~Rangamani and T.~Takayanagi, \emph{{A Covariant holographic
			entanglement entropy proposal}},
	\href{http://dx.doi.org/10.1088/1126-6708/2007/07/062}{\emph{JHEP} {\bf 07}
		(2007) 062}, [\href{https://arxiv.org/abs/0705.0016}{{\tt 0705.0016}}].
	
	\bibitem{Penington:2019kki}
	G.~Penington, S.~H. Shenker, D.~Stanford and Z.~Yang, \emph{{Replica wormholes
			and the black hole interior}},  \href{https://arxiv.org/abs/1911.11977}{{\tt
			1911.11977}}.
	
	\bibitem{Almheiri:2019psy}
	A.~Almheiri, R.~Mahajan and J.~E. Santos, \emph{{Entanglement islands in higher
			dimensions}},
	\href{http://dx.doi.org/10.21468/SciPostPhys.9.1.001}{\emph{SciPost Phys.}
		{\bf 9} (2020) 001}, [\href{https://arxiv.org/abs/1911.09666}{{\tt
			1911.09666}}].
	
	\bibitem{Hashimoto:2020cas}
	K.~Hashimoto, N.~Iizuka and Y.~Matsuo, \emph{{Islands in Schwarzschild black
			holes}}, \href{http://dx.doi.org/10.1007/JHEP06(2020)085}{\emph{JHEP} {\bf
			06} (2020) 085}, [\href{https://arxiv.org/abs/2004.05863}{{\tt 2004.05863}}].
	
	\bibitem{Anegawa:2020ezn}
	T.~Anegawa and N.~Iizuka, \emph{{Notes on islands in asymptotically flat 2d
			dilaton black holes}},
	\href{http://dx.doi.org/10.1007/JHEP07(2020)036}{\emph{JHEP} {\bf 07} (2020)
		036}, [\href{https://arxiv.org/abs/2004.01601}{{\tt 2004.01601}}].
	
	\bibitem{Alishahiha:2020qza}
	M.~Alishahiha, A.~Faraji~Astaneh and A.~Naseh, \emph{{Island in the presence of
			higher derivative terms}},
	\href{http://dx.doi.org/10.1007/JHEP02(2021)035}{\emph{JHEP} {\bf 02} (2021)
		035}, [\href{https://arxiv.org/abs/2005.08715}{{\tt 2005.08715}}].
	
	\bibitem{Gautason:2020tmk}
	F.~F. Gautason, L.~Schneiderbauer, W.~Sybesma and L.~Thorlacius, \emph{{Page
			Curve for an Evaporating Black Hole}},
	\href{http://dx.doi.org/10.1007/JHEP05(2020)091}{\emph{JHEP} {\bf 05} (2020)
		091}, [\href{https://arxiv.org/abs/2004.00598}{{\tt 2004.00598}}].
	
	\bibitem{Hartman:2020swn}
	T.~Hartman, E.~Shaghoulian and A.~Strominger, \emph{{Islands in Asymptotically
			Flat 2D Gravity}},
	\href{http://dx.doi.org/10.1007/JHEP07(2020)022}{\emph{JHEP} {\bf 07} (2020)
		022}, [\href{https://arxiv.org/abs/2004.13857}{{\tt 2004.13857}}].
	
	\bibitem{Hollowood:2020cou}
	T.~J. Hollowood and S.~P. Kumar, \emph{{Islands and Page Curves for Evaporating
			Black Holes in JT Gravity}},
	\href{http://dx.doi.org/10.1007/JHEP08(2020)094}{\emph{JHEP} {\bf 08} (2020)
		094}, [\href{https://arxiv.org/abs/2004.14944}{{\tt 2004.14944}}].
	
	\bibitem{Geng:2020fxl}
	H.~Geng, A.~Karch, C.~Perez-Pardavila, S.~Raju, L.~Randall, M.~Riojas et~al.,
	\emph{{Information Transfer with a Gravitating Bath}},
	\href{http://dx.doi.org/10.21468/SciPostPhys.10.5.103}{\emph{SciPost Phys.}
		{\bf 10} (2021) 103}, [\href{https://arxiv.org/abs/2012.04671}{{\tt
			2012.04671}}].
	
	\bibitem{Geng:2020qvw}
	H.~Geng and A.~Karch, \emph{{Massive islands}},
	\href{http://dx.doi.org/10.1007/JHEP09(2020)121}{\emph{JHEP} {\bf 09} (2020)
		121}, [\href{https://arxiv.org/abs/2006.02438}{{\tt 2006.02438}}].
	
	\bibitem{Li:2020ceg}
	T.~Li, J.~Chu and Y.~Zhou, \emph{{Reflected Entropy for an Evaporating Black
			Hole}}, \href{http://dx.doi.org/10.1007/JHEP11(2020)155}{\emph{JHEP} {\bf 11}
		(2020) 155}, [\href{https://arxiv.org/abs/2006.10846}{{\tt 2006.10846}}].
	
	\bibitem{Chandrasekaran:2020qtn}
	V.~Chandrasekaran, M.~Miyaji and P.~Rath, \emph{{Including contributions from
			entanglement islands to the reflected entropy}},
	\href{http://dx.doi.org/10.1103/PhysRevD.102.086009}{\emph{Phys. Rev. D} {\bf
			102} (2020) 086009}, [\href{https://arxiv.org/abs/2006.10754}{{\tt
			2006.10754}}].
	
	\bibitem{Akers:2019nfi}
	C.~Akers, N.~Engelhardt and D.~Harlow, \emph{{Simple holographic models of
			black hole evaporation}},
	\href{http://dx.doi.org/10.1007/JHEP08(2020)032}{\emph{JHEP} {\bf 08} (2020)
		032}, [\href{https://arxiv.org/abs/1910.00972}{{\tt 1910.00972}}].
	
	\bibitem{Balasubramanian:2020hfs}
	V.~Balasubramanian, A.~Kar, O.~Parrikar, G.~S\'arosi and T.~Ugajin,
	\emph{{Geometric secret sharing in a model of Hawking radiation}},
	\href{http://dx.doi.org/10.1007/JHEP01(2021)177}{\emph{JHEP} {\bf 01} (2021)
		177}, [\href{https://arxiv.org/abs/2003.05448}{{\tt 2003.05448}}].
	
	\bibitem{Hartman:2020khs}
	T.~Hartman, Y.~Jiang and E.~Shaghoulian, \emph{{Islands in cosmology}},
	\href{http://dx.doi.org/10.1007/JHEP11(2020)111}{\emph{JHEP} {\bf 11} (2020)
		111}, [\href{https://arxiv.org/abs/2008.01022}{{\tt 2008.01022}}].
	
	\bibitem{Balasubramanian:2020coy}
	V.~Balasubramanian, A.~Kar and T.~Ugajin, \emph{{Entanglement between two
			disjoint universes}},
	\href{http://dx.doi.org/10.1007/JHEP02(2021)136}{\emph{JHEP} {\bf 02} (2021)
		136}, [\href{https://arxiv.org/abs/2008.05274}{{\tt 2008.05274}}].
	
	\bibitem{Akal:2020twv}
	I.~Akal, Y.~Kusuki, N.~Shiba, T.~Takayanagi and Z.~Wei, \emph{{Entanglement
			Entropy in a Holographic Moving Mirror and the Page Curve}},
	\href{http://dx.doi.org/10.1103/PhysRevLett.126.061604}{\emph{Phys. Rev.
			Lett.} {\bf 126} (2021) 061604},
	[\href{https://arxiv.org/abs/2011.12005}{{\tt 2011.12005}}].
	
	\bibitem{Akal:2021foz}
	I.~Akal, Y.~Kusuki, N.~Shiba, T.~Takayanagi and Z.~Wei, \emph{{Holographic
			moving mirrors}},  \href{https://arxiv.org/abs/2106.11179}{{\tt 2106.11179}}.
	
	\bibitem{Kawabata:2021hac}
	K.~Kawabata, T.~Nishioka, Y.~Okuyama and K.~Watanabe, \emph{{Probing Hawking
			radiation through capacity of entanglement}},
	\href{http://dx.doi.org/10.1007/JHEP05(2021)062}{\emph{JHEP} {\bf 05} (2021)
		062}, [\href{https://arxiv.org/abs/2102.02425}{{\tt 2102.02425}}].
	
	\bibitem{Kawabata:2021vyo}
	K.~Kawabata, T.~Nishioka, Y.~Okuyama and K.~Watanabe, \emph{{Replica wormholes
			and capacity of entanglement}},  \href{https://arxiv.org/abs/2105.08396}{{\tt
			2105.08396}}.
	
	\bibitem{Ling:2020laa}
	Y.~Ling, Y.~Liu and Z.-Y. Xian, \emph{{Island in Charged Black Holes}},
	\href{http://dx.doi.org/10.1007/JHEP03(2021)251}{\emph{JHEP} {\bf 03} (2021)
		251}, [\href{https://arxiv.org/abs/2010.00037}{{\tt 2010.00037}}].
	
	\bibitem{Chen:2020jvn}
	H.~Z. Chen, Z.~Fisher, J.~Hernandez, R.~C. Myers and S.-M. Ruan,
	\emph{{Evaporating Black Holes Coupled to a Thermal Bath}},
	\href{http://dx.doi.org/10.1007/JHEP01(2021)065}{\emph{JHEP} {\bf 01} (2021)
		065}, [\href{https://arxiv.org/abs/2007.11658}{{\tt 2007.11658}}].
	
	\bibitem{Chen:2019uhq}
	H.~Z. Chen, Z.~Fisher, J.~Hernandez, R.~C. Myers and S.-M. Ruan,
	\emph{{Information Flow in Black Hole Evaporation}},
	\href{http://dx.doi.org/10.1007/JHEP03(2020)152}{\emph{JHEP} {\bf 03} (2020)
		152}, [\href{https://arxiv.org/abs/1911.03402}{{\tt 1911.03402}}].
	
	\bibitem{Manu:2020tty}
	A.~Manu, K.~Narayan and P.~Paul, \emph{{Cosmological singularities,
			entanglement and quantum extremal surfaces}},
	\href{http://dx.doi.org/10.1007/JHEP04(2021)200}{\emph{JHEP} {\bf 04} (2021)
		200}, [\href{https://arxiv.org/abs/2012.07351}{{\tt 2012.07351}}].
	
	\bibitem{Chen:2020uac}
	H.~Z. Chen, R.~C. Myers, D.~Neuenfeld, I.~A. Reyes and J.~Sandor,
	\emph{{Quantum Extremal Islands Made Easy, Part I: Entanglement on the
			Brane}}, \href{http://dx.doi.org/10.1007/JHEP10(2020)166}{\emph{JHEP} {\bf
			10} (2020) 166}, [\href{https://arxiv.org/abs/2006.04851}{{\tt 2006.04851}}].
	
	\bibitem{Chen:2020hmv}
	H.~Z. Chen, R.~C. Myers, D.~Neuenfeld, I.~A. Reyes and J.~Sandor,
	\emph{{Quantum Extremal Islands Made Easy, Part II: Black Holes on the
			Brane}}, \href{http://dx.doi.org/10.1007/JHEP12(2020)025}{\emph{JHEP} {\bf
			12} (2020) 025}, [\href{https://arxiv.org/abs/2010.00018}{{\tt 2010.00018}}].
	
	\bibitem{Hollowood:2021nlo}
	T.~J. Hollowood, S.~Prem~Kumar, A.~Legramandi and N.~Talwar, \emph{{Islands in
			the Stream of Hawking Radiation}},
	\href{https://arxiv.org/abs/2104.00052}{{\tt 2104.00052}}.
	
	\bibitem{Ghosh:2021axl}
	K.~Ghosh and C.~Krishnan, \emph{{Dirichlet baths and the not-so-fine-grained
			Page curve}}, \href{http://dx.doi.org/10.1007/JHEP08(2021)119}{\emph{JHEP}
		{\bf 08} (2021) 119}, [\href{https://arxiv.org/abs/2103.17253}{{\tt
			2103.17253}}].
	
	\bibitem{Chu:2021gdb}
	J.~Chu, F.~Deng and Y.~Zhou, \emph{{Page Curve from Defect Extremal Surface and
			Island in Higher Dimensions}},  \href{https://arxiv.org/abs/2105.09106}{{\tt
			2105.09106}}.
	
	\bibitem{Caceres:2021fuw}
	E.~Caceres, A.~Kundu, A.~K. Patra and S.~Shashi, \emph{{Page Curves and Bath
			Deformations}},  \href{https://arxiv.org/abs/2107.00022}{{\tt 2107.00022}}.
	
	\bibitem{Geng:2021hlu}
	H.~Geng, A.~Karch, C.~Perez-Pardavila, S.~Raju, L.~Randall, M.~Riojas et~al.,
	\emph{{Inconsistency of Islands in Theories with Long-Range Gravity}},
	\href{https://arxiv.org/abs/2107.03390}{{\tt 2107.03390}}.
	
	\bibitem{Ahn:2021chg}
	B.~Ahn, S.-E. Bak, H.-S. Jeong, K.-Y. Kim and Y.-W. Sun, \emph{{Islands in
			charged linear dilaton black holes}},
	\href{https://arxiv.org/abs/2107.07444}{{\tt 2107.07444}}.
	
	\bibitem{Krishnan:2020oun}
	C.~Krishnan, V.~Patil and J.~Pereira, \emph{{Page Curve and the Information
			Paradox in Flat Space}},  \href{https://arxiv.org/abs/2005.02993}{{\tt
			2005.02993}}.
	
	\bibitem{Hollowood:2021wkw}
	T.~J. Hollowood, S.~P. Kumar, A.~Legramandi and N.~Talwar, \emph{{Ephemeral
			Islands, Plunging Quantum Extremal Surfaces and BCFT channels}},
	\href{https://arxiv.org/abs/2109.01895}{{\tt 2109.01895}}.
	
	\bibitem{Li:2021dmf}
	T.~Li, M.-K. Yuan and Y.~Zhou, \emph{{Defect Extremal Surface for Reflected
			Entropy}},  \href{https://arxiv.org/abs/2108.08544}{{\tt 2108.08544}}.
	
	\bibitem{Chen:2019iro}
	Y.~Chen, \emph{{Pulling Out the Island with Modular Flow}},
	\href{http://dx.doi.org/10.1007/JHEP03(2020)033}{\emph{JHEP} {\bf 03} (2020)
		033}, [\href{https://arxiv.org/abs/1912.02210}{{\tt 1912.02210}}].
	
	\bibitem{Liu:2020gnp}
	H.~Liu and S.~Vardhan, \emph{{A dynamical mechanism for the Page curve from
			quantum chaos}}, \href{http://dx.doi.org/10.1007/JHEP03(2021)088}{\emph{JHEP}
		{\bf 03} (2021) 088}, [\href{https://arxiv.org/abs/2002.05734}{{\tt
			2002.05734}}].
	
	\bibitem{Akal:2020ujg}
	I.~Akal, \emph{{Universality, intertwiners and black hole information}},
	\href{https://arxiv.org/abs/2010.12565}{{\tt 2010.12565}}.
	
	\bibitem{Krishnan:2020fer}
	C.~Krishnan, \emph{{Critical Islands}},
	\href{http://dx.doi.org/10.1007/JHEP01(2021)179}{\emph{JHEP} {\bf 01} (2021)
		179}, [\href{https://arxiv.org/abs/2007.06551}{{\tt 2007.06551}}].
	
	\bibitem{Krishnan:2021faa}
	C.~Krishnan and V.~Mohan, \emph{{Hints of gravitational ergodicity:
			Berry\textquoteright{}s ensemble and the universality of the semi-classical
			Page curve}}, \href{http://dx.doi.org/10.1007/JHEP05(2021)126}{\emph{JHEP}
		{\bf 05} (2021) 126}, [\href{https://arxiv.org/abs/2102.07703}{{\tt
			2102.07703}}].
	
	\bibitem{Bhattacharya:2020ymw}
	A.~Bhattacharya, \emph{{Multipartite purification, multiboundary wormholes, and
			islands in $AdS_3/CFT_2$}},
	\href{http://dx.doi.org/10.1103/PhysRevD.102.046013}{\emph{Phys. Rev. D} {\bf
			102} (2020) 046013}, [\href{https://arxiv.org/abs/2003.11870}{{\tt
			2003.11870}}].
	
	\bibitem{Basak:2020aaa}
	J.~Kumar~Basak, D.~Basu, V.~Malvimat, H.~Parihar and G.~Sengupta,
	\emph{{Islands for Entanglement Negativity}},
	\href{https://arxiv.org/abs/2012.03983}{{\tt 2012.03983}}.
	
	\bibitem{KumarBasak:2021rrx}
	J.~Kumar~Basak, D.~Basu, V.~Malvimat, H.~Parihar and G.~Sengupta, \emph{{Page
			Curve for Entanglement Negativity through Geometric Evaporation}},
	\href{https://arxiv.org/abs/2106.12593}{{\tt 2106.12593}}.
	
	\bibitem{Geng:2021wcq}
	H.~Geng, Y.~Nomura and H.-Y. Sun, \emph{{Information paradox and its resolution
			in de Sitter holography}},
	\href{http://dx.doi.org/10.1103/PhysRevD.103.126004}{\emph{Phys. Rev. D} {\bf
			103} (2021) 126004}, [\href{https://arxiv.org/abs/2103.07477}{{\tt
			2103.07477}}].
	
	\bibitem{Geng:2021iyq}
	H.~Geng, S.~L\"ust, R.~K. Mishra and D.~Wakeham, \emph{{Holographic BCFTs and
			Communicating Black Holes}},
	\href{http://dx.doi.org/10.1007/JHEP08(2021)003}{\emph{jhep} {\bf 08} (2021)
		003}, [\href{https://arxiv.org/abs/2104.07039}{{\tt 2104.07039}}].
	
	\bibitem{Anderson:2021vof}
	L.~Anderson, O.~Parrikar and R.~M. Soni, \emph{{Islands with Gravitating
			Baths}},  \href{https://arxiv.org/abs/2103.14746}{{\tt 2103.14746}}.
	
	\bibitem{Caceres:2020jcn}
	E.~Caceres, A.~Kundu, A.~K. Patra and S.~Shashi, \emph{{Warped Information and
			Entanglement Islands in AdS/WCFT}},
	\href{http://dx.doi.org/10.1007/JHEP07(2021)004}{\emph{JHEP} {\bf 07} (2021)
		004}, [\href{https://arxiv.org/abs/2012.05425}{{\tt 2012.05425}}].
	
	\bibitem{Reyes:2021npy}
	I.~A. Reyes, \emph{{Moving Mirrors, Page Curves, and Bulk Entropies in AdS2}},
	\href{http://dx.doi.org/10.1103/PhysRevLett.127.051602}{\emph{Phys. Rev.
			Lett.} {\bf 127} (2021) 051602},
	[\href{https://arxiv.org/abs/2103.01230}{{\tt 2103.01230}}].
	
	\bibitem{Neuenfeld:2021bsb}
	D.~Neuenfeld, \emph{{Double Holography as a Model for Black Hole
			Complementarity}},  \href{https://arxiv.org/abs/2105.01130}{{\tt
			2105.01130}}.
	
	\bibitem{Matsuo:2020ypv}
	Y.~Matsuo, \emph{{Islands and stretched horizon}},
	\href{http://dx.doi.org/10.1007/JHEP07(2021)051}{\emph{JHEP} {\bf 07} (2021)
		051}, [\href{https://arxiv.org/abs/2011.08814}{{\tt 2011.08814}}].
	
	\bibitem{Azarnia:2021uch}
	S.~Azarnia, R.~Fareghbal, A.~Naseh and H.~Zolfi, \emph{{Islands in Flat-Space
			Cosmology}},  \href{https://arxiv.org/abs/2109.04795}{{\tt 2109.04795}}.
	
	\bibitem{Miyaji:2021lcq}
	M.~Miyaji, \emph{{Entanglement of Initial State and Pseudo Entanglement
			Wedge}},  \href{https://arxiv.org/abs/2109.03830}{{\tt 2109.03830}}.
	
	\bibitem{Almheiri:2020cfm}
	A.~Almheiri, T.~Hartman, J.~Maldacena, E.~Shaghoulian and A.~Tajdini,
	\emph{{The entropy of Hawking radiation}},
	\href{http://dx.doi.org/10.1103/RevModPhys.93.035002}{\emph{Rev. Mod. Phys.}
		{\bf 93} (2021) 035002}, [\href{https://arxiv.org/abs/2006.06872}{{\tt
			2006.06872}}].
	
	\bibitem{Bhattacharya:2020uun}
	A.~Bhattacharya, A.~Chanda, S.~Maulik, C.~Northe and S.~Roy, \emph{{Topological
			shadows and complexity of islands in multiboundary wormholes}},
	\href{http://dx.doi.org/10.1007/JHEP02(2021)152}{\emph{JHEP} {\bf 02} (2021)
		152}, [\href{https://arxiv.org/abs/2010.04134}{{\tt 2010.04134}}].
	
	\bibitem{Nielsen1133}
	M.~A. Nielsen, M.~R. Dowling, M.~Gu and A.~C. Doherty, \emph{Quantum
		computation as geometry},
	\href{http://dx.doi.org/10.1126/science.1121541}{\emph{Science} {\bf 311}
		(2006) 1133--1135}, [\href{https://arxiv.org/abs/quant-ph/0603161}{{\tt
			quant-ph/0603161}}].
	
	\bibitem{Nielsen2006optimal}
	M.~A. Nielsen, M.~R. Dowling, M.~Gu and A.~C. Doherty, \emph{Optimal control,
		geometry, and quantum computing},
	\href{http://dx.doi.org/10.1103/physreva.73.062323}{\emph{Physical Review A}
		{\bf 73} (Jun, 2006) }.
	
	\bibitem{Bhattacharya:2021jrn}
	A.~Bhattacharya, A.~Bhattacharyya, P.~Nandy and A.~K. Patra, \emph{{Islands and
			complexity of eternal black hole and radiation subsystems for a doubly
			holographic model}},
	\href{http://dx.doi.org/10.1007/JHEP05(2021)135}{\emph{JHEP} {\bf 05} (2021)
		135}, [\href{https://arxiv.org/abs/2103.15852}{{\tt 2103.15852}}].
	
	\bibitem{Maldacena:2013xja}
	J.~Maldacena and L.~Susskind, \emph{{Cool horizons for entangled black holes}},
	\href{http://dx.doi.org/10.1002/prop.201300020}{\emph{Fortsch. Phys.} {\bf
			61} (2013) 781--811}, [\href{https://arxiv.org/abs/1306.0533}{{\tt
			1306.0533}}].
	
	\bibitem{Verlinde:2020upt}
	H.~Verlinde, \emph{{ER = EPR revisited: On the Entropy of an Einstein-Rosen
			Bridge}},  \href{https://arxiv.org/abs/2003.13117}{{\tt 2003.13117}}.
	
	\bibitem{Israel:1966rt}
	W.~Israel, \emph{{Singular hypersurfaces and thin shells in general
			relativity}}, \href{http://dx.doi.org/10.1007/BF02710419}{\emph{Nuovo Cim. B}
		{\bf 44S10} (1966) 1}.
	
	\bibitem{Caceres:2019giy}
	E.~Caceres, A.~Kundu, A.~K. Patra and S.~Shashi, \emph{{A Killing Vector
			Treatment of Multiboundary Wormholes}},
	\href{http://dx.doi.org/10.1007/JHEP02(2020)149}{\emph{JHEP} {\bf 02} (2020)
		149}, [\href{https://arxiv.org/abs/1912.08793}{{\tt 1912.08793}}].
	
	\bibitem{Bhattacharyya:2016hbx}
	A.~Bhattacharyya, Z.-S. Gao, L.-Y. Hung and S.-N. Liu, \emph{{Exploring the
			Tensor Networks/AdS Correspondence}},
	\href{http://dx.doi.org/10.1007/JHEP08(2016)086}{\emph{JHEP} {\bf 08} (2016)
		086}, [\href{https://arxiv.org/abs/1606.00621}{{\tt 1606.00621}}].
	
	\bibitem{Balasubramanian:2014hda}
	V.~Balasubramanian, P.~Hayden, A.~Maloney, D.~Marolf and S.~F. Ross,
	\emph{{Multiboundary Wormholes and Holographic Entanglement}},
	\href{http://dx.doi.org/10.1088/0264-9381/31/18/185015}{\emph{Class. Quant.
			Grav.} {\bf 31} (2014) 185015}, [\href{https://arxiv.org/abs/1406.2663}{{\tt
			1406.2663}}].
	
	\bibitem{Susskind:2014moa}
	L.~Susskind, \emph{{Entanglement is not enough}},
	\href{http://dx.doi.org/10.1002/prop.201500095}{\emph{Fortsch. Phys.} {\bf
			64} (2016) 49--71}, [\href{https://arxiv.org/abs/1411.0690}{{\tt
			1411.0690}}].
	
	\bibitem{Susskind:2014rva}
	L.~Susskind, \emph{{Computational Complexity and Black Hole Horizons}},
	\href{http://dx.doi.org/10.1002/prop.201500092}{\emph{Fortsch. Phys.} {\bf
			64} (2016) 24--43}, [\href{https://arxiv.org/abs/1403.5695}{{\tt
			1403.5695}}].
	
	\bibitem{Agon:2018zso}
	C.~A. Ag\'on, M.~Headrick and B.~Swingle, \emph{{Subsystem Complexity and
			Holography}}, \href{http://dx.doi.org/10.1007/JHEP02(2019)145}{\emph{JHEP}
		{\bf 02} (2019) 145}, [\href{https://arxiv.org/abs/1804.01561}{{\tt
			1804.01561}}].
	
	\bibitem{Alishahiha:2018lfv}
	M.~Alishahiha, K.~Babaei~Velni and M.~R. Mohammadi~Mozaffar, \emph{{Black hole
			subregion action and complexity}},
	\href{http://dx.doi.org/10.1103/PhysRevD.99.126016}{\emph{Phys. Rev. D} {\bf
			99} (2019) 126016}, [\href{https://arxiv.org/abs/1809.06031}{{\tt
			1809.06031}}].
	
	\bibitem{Chen:2018mcc}
	B.~Chen, W.-M. Li, R.-Q. Yang, C.-Y. Zhang and S.-J. Zhang, \emph{{Holographic
			subregion complexity under a thermal quench}},
	\href{http://dx.doi.org/10.1007/JHEP07(2018)034}{\emph{JHEP} {\bf 07} (2018)
		034}, [\href{https://arxiv.org/abs/1803.06680}{{\tt 1803.06680}}].
	
	\bibitem{Alishahiha:2015rta}
	M.~Alishahiha, \emph{{Holographic Complexity}},
	\href{http://dx.doi.org/10.1103/PhysRevD.92.126009}{\emph{Phys. Rev. D} {\bf
			92} (2015) 126009}, [\href{https://arxiv.org/abs/1509.06614}{{\tt
			1509.06614}}].
	
	\bibitem{Abt:2017pmf}
	R.~Abt, J.~Erdmenger, H.~Hinrichsen, C.~M. Melby-Thompson, R.~Meyer, C.~Northe
	et~al., \emph{{Topological Complexity in AdS$_3$/CFT$_2$}},
	\href{http://dx.doi.org/10.1002/prop.201800034}{\emph{Fortsch. Phys.} {\bf
			66} (2018) 1800034}, [\href{https://arxiv.org/abs/1710.01327}{{\tt
			1710.01327}}].
	
	\bibitem{Abt:2018ywl}
	R.~Abt, J.~Erdmenger, M.~Gerbershagen, C.~M. Melby-Thompson and C.~Northe,
	\emph{{Holographic Subregion Complexity from Kinematic Space}},
	\href{http://dx.doi.org/10.1007/JHEP01(2019)012}{\emph{JHEP} {\bf 01} (2019)
		012}, [\href{https://arxiv.org/abs/1805.10298}{{\tt 1805.10298}}].
	
	\bibitem{Bhattacharya:2019zkb}
	A.~Bhattacharya, K.~T. Grosvenor and S.~Roy, \emph{{Entanglement Entropy and
			Subregion Complexity in Thermal Perturbations around Pure-AdS Spacetime}},
	\href{http://dx.doi.org/10.1103/PhysRevD.100.126004}{\emph{Phys. Rev. D} {\bf
			100} (2019) 126004}, [\href{https://arxiv.org/abs/1905.02220}{{\tt
			1905.02220}}].
	
	\bibitem{Auzzi:2021nrj}
	R.~Auzzi, S.~Baiguera, S.~Bonansea, G.~Nardelli and K.~Toccacelo, \emph{{Volume
			complexity for Janus $\mathrm{AdS}_3$ geometries}},
	\href{http://dx.doi.org/10.1007/JHEP08(2021)045}{\emph{JHEP} {\bf 08} (2021)
		045}, [\href{https://arxiv.org/abs/2105.08729}{{\tt 2105.08729}}].
	
	\bibitem{Baiguera:2021cba}
	S.~Baiguera, S.~Bonansea and K.~Toccacelo, \emph{{Volume complexity for the
			non-supersymmetric Janus AdS$_5$ geometry}},
	\href{https://arxiv.org/abs/2105.12743}{{\tt 2105.12743}}.
	
	\bibitem{Sato:2021ftf}
	Y.~Sato, \emph{{Complexity in a moving mirror model}},
	\href{https://arxiv.org/abs/2108.04637}{{\tt 2108.04637}}.
	
	\bibitem{Bernamonti:2021jyu}
	A.~Bernamonti, F.~Bigazzi, D.~Billo, L.~Faggi and F.~Galli, \emph{{Holographic
			and QFT Complexity with angular momentum}},
	\href{https://arxiv.org/abs/2108.09281}{{\tt 2108.09281}}.
	
	\bibitem{Hernandez:2020nem}
	J.~Hernandez, R.~C. Myers and S.-M. Ruan, \emph{{Quantum extremal islands made
			easy. Part III. Complexity on the brane}},
	\href{http://dx.doi.org/10.1007/JHEP02(2021)173}{\emph{JHEP} {\bf 02} (2021)
		173}, [\href{https://arxiv.org/abs/2010.16398}{{\tt 2010.16398}}].
	
	\bibitem{Hayden:2021gno}
	P.~Hayden, O.~Parrikar and J.~Sorce, \emph{{The Markov gap for geometric
			reflected entropy}},  \href{https://arxiv.org/abs/2107.00009}{{\tt
			2107.00009}}.
	
	\bibitem{Miyaji:2015woj}
	M.~Miyaji, T.~Numasawa, N.~Shiba, T.~Takayanagi and K.~Watanabe,
	\emph{{Distance between Quantum States and Gauge-Gravity Duality}},
	\href{http://dx.doi.org/10.1103/PhysRevLett.115.261602}{\emph{Phys. Rev.
			Lett.} {\bf 115} (2015) 261602},
	[\href{https://arxiv.org/abs/1507.07555}{{\tt 1507.07555}}].
	
	\bibitem{Alishahiha:2017cuk}
	M.~Alishahiha and A.~Faraji~Astaneh, \emph{{Holographic Fidelity
			Susceptibility}},
	\href{http://dx.doi.org/10.1103/PhysRevD.96.086004}{\emph{Phys. Rev. D} {\bf
			96} (2017) 086004}, [\href{https://arxiv.org/abs/1705.01834}{{\tt
			1705.01834}}].
	
	\bibitem{Flory:2017ftd}
	M.~Flory, \emph{{A complexity/fidelity susceptibility $g$-theorem for
			AdS$_{3}$/BCFT$_{2}$}},
	\href{http://dx.doi.org/10.1007/JHEP06(2017)131}{\emph{JHEP} {\bf 06} (2017)
		131}, [\href{https://arxiv.org/abs/1702.06386}{{\tt 1702.06386}}].
	
	\bibitem{Banerjee:2017qti}
	S.~Banerjee, J.~Erdmenger and D.~Sarkar, \emph{{Connecting Fisher information
			to bulk entanglement in holography}},
	\href{http://dx.doi.org/10.1007/JHEP08(2018)001}{\emph{JHEP} {\bf 08} (2018)
		001}, [\href{https://arxiv.org/abs/1701.02319}{{\tt 1701.02319}}].
	
	\bibitem{Freedman:2016zud}
	M.~Freedman and M.~Headrick, \emph{{Bit threads and holographic entanglement}},
	\href{http://dx.doi.org/10.1007/s00220-016-2796-3}{\emph{Commun. Math. Phys.}
		{\bf 352} (2017) 407--438}, [\href{https://arxiv.org/abs/1604.00354}{{\tt
			1604.00354}}].
	
	\bibitem{Pedraza:2021mkh}
	J.~F. Pedraza, A.~Russo, A.~Svesko and Z.~Weller-Davies, \emph{{Lorentzian
			threads as 'gatelines' and holographic complexity}},
	\href{https://arxiv.org/abs/2105.12735}{{\tt 2105.12735}}.
	
	\bibitem{Pedraza:2021fgp}
	J.~F. Pedraza, A.~Russo, A.~Svesko and Z.~Weller-Davies, \emph{{Sewing
			spacetime with Lorentzian threads: complexity and the emergence of time in
			quantum gravity}},  \href{https://arxiv.org/abs/2106.12585}{{\tt
			2106.12585}}.
	
\end{thebibliography}
\providecommand{\href}[2]{#2}\begingroup\raggedright\endgroup

\end{document}